\documentclass[12pt]{article}
\usepackage{bezier}
\usepackage{hyperref}
\usepackage{amssymb}
\usepackage{amsbsy}
\usepackage{amsmath}
\usepackage{url}

\def\baselinestretch{1.1}
\renewcommand{\thefootnote}{\fnsymbol{footnote}}
\newcommand{\newsection}{    
\setcounter{equation}{0}
\section}
\def\appendix#1{
  \addtocounter{section}{1}
  \setcounter{equation}{0}
  \renewcommand{\thesection}{\Alph{section}}
  \section*{Appendix \thesection\protect\indent #1}
  \addcontentsline{toc}{section}{Appendix \thesection\ \ \ #1}
  }
\newcommand{\tr}[1]{\,{\rm tr}#1}
\newcommand{\ntr}[1]{\,\frac {\rm tr}{N}#1}
\def\e{{\,\rm e}\,}

\def\eop{\vspace*{\fill}\pagebreak}
\def\be{\begin{equation}}
\def\ee{\end{equation}}
\def\bea{\begin{eqnarray}}
\def\eea{\end{eqnarray}}
\def\LA{\left\langle}
\def\RA{\right\rangle}
\def\BLA{\Big\langle}
\def\BRA{\Big\rangle}

\newcommand{\rf}[1]{(\ref{#1})}
\newcommand{\eq}[1]{Eq.~(\ref{#1})}

\def\a{\alpha}

\def\d{\partial}
\def\L{\Lambda}
\def\l{\lambda}
\def\h{\eta}
\def\om{\omega}
\def\s{\sigma}
\def\P{\hbox{P}}
\newcommand{\bbox}[1]{\boldsymbol{#1}}
\newcommand{\ie}{{\it i.e.}\ }
\newcommand{\half}{{\textstyle{1\over 2}}}
\newcommand{\p}{{\prime}}
\newcommand{\ra}{\rightarrow}
\newcommand{\fr}[2]{{\textstyle {#1 \over #2}}}

\newcommand{\pintc}{\int_C \hspace{-1.31em}\not\hspace{0.85em}}
\newcommand{\pintab}{\int_0^1 \hspace{-1.58em}\not\hspace{1.10em}}
\newcommand{\pint}{\int\hspace{-1.15em}\not\hspace{0.7em}}
\newcommand{\eps}{\varepsilon}

\newcommand{\non}{\nonumber \\*}

\newcommand{\re}{\,\hbox{Re}\,}
\newcommand{\im}{\,\hbox{Im}\,}
\def\la{\mathrel{\mathpalette\fun <}}

\def\fun#1#2{\lower3.6pt\vbox{\baselineskip0pt\lineskip.9pt
\ialign{$\mathsurround=0pt#1\hfil##\hfil$\crcr#2\crcr\sim\crcr}}}

\begin{document}

\begin{titlepage}
\begin{flushright}
YM-7-94 \\ 
December, 1994 \\
\end{flushright}
\vspace{.5cm}

\begin{center}                                               
{\LARGE Notes on the Loop Equation in Loop Space}  
\end{center} \vspace{1cm}
\begin{center}
{\large Yuri Makeenko}\footnote{~E--mail: \ makeenko@nbi.dk \ / \
makeenko@itep.ru \ }
\\ \mbox{} \\
{\it The Niels Bohr Institute,} \\
{\it Blegdamsvej 17, 2100 Copenhagen, DK} \\ \vskip .2 cm
and  \\  \vskip .2 cm
{\it Institute of Theoretical and Experimental Physics,} 
\\ {\it B. Cheremushkinskaya 25, 117259 Moscow, RF} 
\end{center}

\vskip 1 cm
\begin{abstract}
The loop equation satisfied by Wilson's loops in QCD is reformulated as a functional
Laplace equation.  Discretizing the loop space by polygons, Green's function of the functional
 Laplacian is represented as a path integral of the Euclidean harmonic oscillator and is applied for 
 an iterative solution of the equation. It is shown how the usual Feynman's diagrams are
  reproduced through  order $(g^2N)^2$ including the one with the three-gluon vertex.
 
\end{abstract}
                         
\vspace{1cm}
\noindent

\eop
\end{titlepage}
\setcounter{page}{2}
\renewcommand{\thefootnote}{\arabic{footnote}}
\setcounter{footnote}{0}

\renewcommand{\baselinestretch}{0.8}
\small
\tableofcontents
\eop
\renewcommand{\baselinestretch}{1.1}
\normalsize
\setcounter{page}{3}

\newsection{Introduction}

These notes were written in the late 1980s -- the early 1990s and deals with the reformulation of the original loop equation~\cite{MM79,MM81} (see~\cite{Mig83} for a review)
entirely in the loop space  whose elements are arbitrary closed loops. 
The motivation was to develop a scheme for the perturbative 
(and potentially nonperturbative) solution of the loop equation 
as a functional Laplace equation on the loop space by iterations
and, more generally, to generalize the mathematical technique described
 in the books~\cite{Lev51,Fel86} to the functionals of
the type of Wilson's loops which are specific to Gauge Theory.
The parts of the notes are used in the book~\cite{Mak02} and in several lectures/reviews 
but was never published as a whole.
A described new result is about the emergence of the three-gluon vertex in the process of
the iterative solution by inverting the loop-space Laplacian.
The decision to publish the notes now is caused by the very interesting recent advances in solving the loop equations both on the lattice\cite{AK17,KZ23,Li24,KZ24,Boo25} and 
in the momentum-loop space~\cite{Mig25} including turbulence~\cite{turb} 
as well as by several papers on mathematical aspects of the loop equations postered in 
arXiv on [math-ph] and [math.PR].
I have refrained from extensive editing the old text -- only have included some more recent references.

Sect.~\ref{s1} is devoted to the reformulation of the original (vector) loop equation as a (scalar) functional Laplace equation by integrating the former over the loop. The latter is well-defined for a wider class of functionals than the phase factors.
 In Sect.~\ref{s2} the loop space is discretized, approximating loops by polygons. The key attention is paid to the continuity as is required for the lines of force in a gauge theory. 
The corresponding discterization of the functional
 Laplacian is constructed together with its Green's function.
 The loop-space Laplacian is approximated
by a second-order partial differential operator of a specific form. 
Its Green's  function is
represented in the form of a Gaussian integral whose continuum limit takes 
the form of a path integral for the Euclidean harmonic oscillator at finite temperature.
In Sect.~\ref{s3} the continuum Green function is alternatively derived introducing a
smearing in the functional Laplacian with the smearing  parameter $\eps\to 0$.
The rules for calculations of the associated averages are developed and used in Sect.~\ref{s4}
for an iterative solution of the loop equation, where it is shown how the three-gluon vertex
emerges in a nontrivial way by doing the uncertainty $\eps\times \eps^{-1}$.

\newsection{Loop equation as a functional Laplace equation \label{s1}}

The loop-space approach in QCD consists in reformulating QCD entirely in terms 
of gauge-invariant (colorless) objects --- the Wilson loops.  It is based on 
the possibility ($1$) to express observables via the Wilson loops and ($2$) to 
rewrite dynamical equations of motion via the Wilson loops. 
In the large-$N$ limit 
these equations reduce to a single closed equation for the vacuum 
expectation value of the Wilson loop which is known as the loop equation.
This equation is functional but
classical in the sense that it does not involve higher 
correlators of the Wilson loops. It
admits several equivalent formulations. \mbox{A very} convenient one 
is based on the representation of the loop equation as the functional Laplace 
equation on loop space. The last form of the loop equation is closely related 
to the method of stochastic quantization which suggests a consistent
nonperturbative regularization. 

\subsection{Loop space}

Loop space consists of arbitrary continuous closed loops, $C$.
They can be described in a parametric form by the functions 
$x_\mu(\sigma)\in L_2$,
where $\sigma_0\leq\sigma\leq\sigma_f$ and $\mu=1,\ldots,d$, 
which take the values 
in a $d$-dimensional Euclidean space.%
\footnote{~Let us remind that $L_2$ stands for the Hilbert space of functions
$x_\mu(\s)$ whose square is integrable over the Lebesgue measure: 
$\int_{\s_0}^{\s_f} d \s x^2_\mu(\s) < \infty $.}
The functions $x_\mu(\sigma)$ can be 
discontinuous, generally speaking, for an arbitrary choice of the parameter 
$\s$. The continuity of the loop $C$ implies a continuous dependence 
on parameters of the type of proper length
\be
s(\s) = \int_{\s_0}^{\s} d\s^\p \sqrt{\dot{x}^2_\mu(\s^\p)}
\label{properlength}
\ee
where $\dot{x}_\mu(\s)=dx_\mu(\s)/d\s$. 

The functions $x_\mu(\s)\in L_2$ which are associated with the elements of loop 
space obey the following restrictions:
\vspace{-10pt}
\begin{itemize}
\addtolength{\itemsep}{-8pt}
\item[i)] The points $\s=\s_0$ and $\s=\s_f$ are identified:  
$x_\mu(\s_0)=x_\mu(\s_f)$
--- the loops are closed.

\item[ii)] 
The functions $x_\mu(\s)$ and $\Lambda_{\mu\nu}x_\nu(\s)+\alpha_\mu$, with 
$\Lambda_{\mu\nu}$ and $\alpha_\mu$ being independent of $\s$, represent the 
same element of the loop space --- rotational and translational invariance.

\item[iii)] 
The functions $x_\mu(\s)$ and $x_\mu(\s^\p)$ with $\s^\p=f(\s),~f^\p(\s)\geq0$
describe the same loop --- reparametrization invariance.
\end{itemize}
\vspace{-10pt}
Two famous examples of loop space are ($1$) 
the lines of force in Yang--Mills 
theory and ($2$) trajectories of the end of a string.

We shall consider functionals which are defined on the elements of loop space.
The crucial role in Yang--Mills theory is played by the Wilson loops
(the traces of non-Abelian phase factors)
\be
\Phi[A,C] \equiv \ntr{} \Big( \hbox{P} 
\e^{ \int _{C} d\xi_\mu A_\mu(\xi)} \Big)
\label{Wl}
\ee
and their vacuum expectation values
\be
W(C) = \BLA \Phi[A,C] \BRA_A~.
\label{Wilsonloop}
\ee
Here $A_\mu(x)=ig \sum_{a=1}^{N^2-1} t^a A^a_\mu(x)$ is an anti-Hermitean matrix
from the adjoint representation of
the gauge group $SU(N)$ with $t^a$ ($a=1,\ldots,N^2$--$1$)
being the generators of the $SU(N)$. 
The average is defined by
\be
\BLA {\cal F}[A] \BRA_A \equiv \frac
{\int {\cal D} A \e^{\frac {1}{4g^2}\int d^d x \tr{} F_{\mu\nu}^2(x)} 
{\cal F}[A]}
{\int {\cal D} A \e^{\frac {1}{4g^2}\int d^d x \tr{} F_{\mu\nu}^2(x)}}
\ee
with
\be
F_{\mu\nu}(x) = \frac{\d}{\d x_\mu} A_\nu(x) - \frac{\d}{\d x_\nu} A_\mu(x)
+ [A_\mu(x),A_\nu(x)]
\label{fieldstrength}
\ee
being non-Abelian field strength in $d=4$ Euclidean dimensions.

Another example of functionals defined on loop space is a wave functional in 
string theory.

\subsection{Differential calculus on loop space}

The standard form of the loop equation involves derivatives 
w.r.t.\  {\it area}\/ of an infinitesimal loop 
$\delta C_{\mu\nu}(x)$ attached to the given loop at the point $x$ in the 
$\mu\nu$-plane --- $\delta/\delta\s_{\mu\nu}(x)$, and w.r.t.\
{\it path}\/ $\delta x_\mu$                  
along which the point $x$ is shifted from the loop --- $\d^x_\mu$. 

The area derivative is defined by
\vskip .2 cm
\unitlength=.8mm
\message{Be patient -- drawing figures} 
\begin{equation}
\frac{\delta {\cal F}(C)}{ \delta \sigma_{\mu\nu}(x)}  \equiv
\frac{1}{\left| \delta \sigma_{\mu\nu}\right|}
\left[ \;
{\cal F} \left( \hbox{
\begin{picture}(44.0,13.00)(10,112)
\thicklines
\bezier{80}(11.00,108.00)(11.00,119.00)(19.00,119.00)
\bezier{64}(19.00,119.00)(28.00,121.00)(32.00,126.00)
\bezier{108}(32.00,126.00)(39.00,133.00)(41.00,118.00)
\bezier{44}(41.00,118.00)(44.00,122.00)(45.00,118.00)
\bezier{36}(41.00,117.00)(44.50,113.50)(45.00,118.00)
\bezier{148}(21.00,100.00)(43.00,101.00)(41.00,117.00)
\bezier{80}(11.00,108.00)(11.00,100.00)(21.00,100.00)
\put(39.00,117.50){\makebox(0,0)[cc]{$x$}}
\put(47.50,115.00){\vector(1,0){8.00}}
\put(47.50,115.00){\vector(4,3){5.0}}
\put(52.50,111.0){\makebox(0,0)[cb]{$\mu$}}
\put(50.00,118.00){\makebox(0,0)[rb]{$\nu$}}
\end{picture} }
\right)
~-~ {\cal F} \left( \hbox{
\begin{picture}(33.00,13.00)(10,112)
\thicklines
\bezier{80}(11.00,108.00)(11.00,119.00)(19.00,119.00)
\bezier{64}(19.00,119.00)(28.00,121.00)(32.00,126.00)
\bezier{108}(32.00,126.00)(39.00,133.00)(41.00,117.50)
\bezier{148}(21.00,100.00)(43.00,101.00)(41.00,117.50)
\bezier{80}(11.00,108.00)(11.00,100.00)(21.00,100.00)
\put(39.00,117.50){\makebox(0,0)[cc]{$x$}}
\end{picture} } \right) 
\right]
\label{aa}
\end{equation}
\vskip .2 cm \noindent
where ${\left| \delta \sigma_{\mu\nu}\right|}$ stands for the area 
enclosed by the infinitesimal loop in the $\mu\nu$-plane.

Analogously, the path derivative is defined by
\vskip .2 cm
\unitlength=.8mm
\begin{equation}
\partial_\mu^x \, {{\cal F}(C_{xx})}  \equiv
\frac{1}{\left| \delta x_{\mu}\right|}
\left[ \;
{\cal F} \left( \hbox{
\begin{picture}(44.0,13.00)(10,112)
\thicklines
\bezier{80}(11.00,108.00)(11.00,119.00)(19.00,119.00)
\bezier{64}(19.00,119.00)(28.00,121.00)(32.00,126.00)
\bezier{108}(32.00,126.00)(39.00,133.00)(41.00,118.00)
\bezier{148}(21.00,100.00)(43.00,101.00)(41.00,117.00)
\bezier{80}(11.00,108.00)(11.00,100.00)(21.00,100.00)
\put(39.00,117.50){\makebox(0,0)[cc]{$x$}}
\put(47.50,115.00){\vector(1,0){8.00}}
\put(52.50,111.00){\makebox(0,0)[cb]{$\mu$}}
\put(46.20,117.50){\circle*{1.25}}
\bezier{16}(41.00,118.00)(43.50,117.00)(46.00,118.00)
\bezier{16}(41.00,117.00)(43.50,116.00)(46.00,117.00)
\end{picture} }
\right)
~-~ {\cal F} \left( \hbox{
\begin{picture}(33.00,13.00)(10,112)
\thicklines
\bezier{80}(11.00,108.00)(11.00,119.00)(19.00,119.00)
\bezier{64}(19.00,119.00)(28.00,121.00)(32.00,126.00)
\bezier{108}(32.00,126.00)(39.00,133.00)(41.00,117.50)
\bezier{148}(21.00,100.00)(43.00,101.00)(41.00,117.50)
\bezier{80}(11.00,108.00)(11.00,100.00)(21.00,100.00)
\put(38.80,117.50){\makebox(0,0)[cc]{$x$}}
\put(41.20,117.50){\circle*{1.25}}
\end{picture} } \right) 
\right]
\label{pp}
\end{equation}
\vskip .2 cm \noindent
where ${\left| \delta x_{\mu}\right|}$ stands for the length of the 
infinitesimal path.

These two differential
operations are well-defined for so-called functionals of the Stokes type which 
satisfy the backtracking condition --- they do not change when an appendix 
passing back and forth is added to the loop at some point $x$: 
\vskip .2 cm
\unitlength=.8mm
\begin{equation}
{\cal F} \left( \hbox{
\begin{picture}(44.0,13.00)(10,112)
\thicklines
\bezier{80}(11.00,108.00)(11.00,119.00)(19.00,119.00)
\bezier{64}(19.00,119.00)(28.00,121.00)(32.00,126.00)
\bezier{108}(32.00,126.00)(39.00,133.00)(41.00,118.00)
\bezier{148}(21.00,100.00)(43.00,101.00)(41.00,116.00)
\bezier{80}(11.00,108.00)(11.00,100.00)(21.00,100.00)
\bezier{49}(41.00,118.00)(49.00,118.00)(49.00,124.00)
\bezier{49}(41.00,116.00)(51.00,116.00)(51.00,124.00)
\put(50.2,124.00){\oval(2.0,2.0)[t]}
\put(39.00,117.00){\makebox(0,0)[cc]{$x$}}
\end{picture} }
\right)
~=~ {\cal F} \left( \hbox{
\begin{picture}(33.00,13.00)(10,112)
\thicklines
\bezier{80}(11.00,108.00)(11.00,119.00)(19.00,119.00)
\bezier{64}(19.00,119.00)(28.00,121.00)(32.00,126.00)
\bezier{108}(32.00,126.00)(39.00,133.00)(41.00,117.50)
\bezier{148}(21.00,100.00)(43.00,101.00)(41.00,117.50)
\bezier{80}(11.00,108.00)(11.00,100.00)(21.00,100.00)
\end{picture} } \right)~.
\label{btr}
\end{equation}
\vskip .2 cm \noindent
This condition is equivalent to
the Bianchi identity of Yang--Mills theory and is obviously satisfied by 
the Wilson loop~\rf{Wl} due to the properties of the non-Abelian
phase factor. Such functionals are 
known in mathematics as Chen integrals~\cite{Tav93}. 
For the Stokes functionals the variation on the r.h.s.\ of \eq{aa} 
is proportional to the area enclosed by the infinitesimal loop 
$\delta C_{\mu\nu}(x)$ and does not depend on its shape. 
Analogously, the variation on the r.h.s.\ of \eq{pp} 
is proportional to the length of the infinitesimal path $\delta 
x_{\mu}$ and does not depend on its shape. 

If $x$ is a regular point (like any point of the contour for the 
functional~\rf{Wl}), the r.h.s. \ of \eq{pp} vanishes due to the backtracking 
condition~\rf{btr}. In order for the result to be nonvanishing the point $x$
should be a {\it marked}\/ (or irregular) point. 
The simplest example of the functional with a marked point $x$ is 
\be
\Phi^a[A,C_{xx}] \equiv \ntr{} \Big( t^a \, \hbox{P} 
\e^{ \int _{C_{xx}} d\xi_\mu A_\mu(\xi)} \Big)
\label{Wlmarked}
\ee
with the $SU(N)$ generator $t^a$  inserted in the 
path-ordered product at the point $x$.

The standard variational derivative, 
$\delta/\delta x_\mu(\s)$, can be expressed via the path and area derivatives 
by the formula
\be
\frac{\delta}{\delta x_\mu(\s)} = \dot{x}_\nu(\s) 
\frac{\delta}{\delta\s_{\mu\nu}( x(\s) )} +
\sum_{i=1}^m \d_\mu^{x_i} \delta(\s-\s_i)
\label{varder}
\ee
where the sum on the r.h.s.\ is present for the case of a functional having 
$m$ marked (irregular) points $x_i\equiv x(\s_i)$. A simplest example of 
the functional with $m$ marked points is just a function of $m$ variables 
$x_1,\ldots,x_m$. 

Making use of \eq{varder}, the path derivative can be calculated as the 
limiting procedure
\be
\d_\mu^{x(\s)}= \int_{\s-0}^{\s+0} d \s^\p
\frac{\delta}{\delta x_\mu(\s^\p)}  \,.
\label{vpath}
\ee
The result is obviously nonvanishing only when $\d_\mu^x$ is applied to a 
functional with $x(\s)$ being a marked point.

It is nontrivial that the area derivative can also be expressed via
the variational derivative~\cite{Pol80}:
\be
\frac{\delta}{\delta\s_{\mu\nu}( x(\s) )} =
 \int_{\s-0}^{\s+0} d \s^\p (\s^\p -\s)
\frac{\delta}{\delta x_\mu(\s^\p)}  
\frac{\delta}{\delta x_\nu(\s)}  \,.
\label{varea}
\ee
The point is that the six-component quantity ,
${\delta}/{\delta\s_{\mu\nu}( x(\s) )} $,
is expressed via the four-component one,
${\delta}/{\delta x_{\mu}(\s)} $, which is possible because
the components of
${\delta}/{\delta\s_{\mu\nu}( x(\s) )} $ are dependent
due to the Bianchi identity.

\subsection{Loop equation on loop space}

The original form of the loop equation of large-$N$ QCD,
which is written for the vacuum 
expectation value of the Wilson loop~\rf{Wilsonloop}, reads~\cite{MM79,MM81}
\be
\d_\mu^x \frac{\delta}{\delta\s_{\mu\nu}( x)} W(C) =
\lambda \pintc d x_\mu^\p \delta^{(d)}(x-x^\p) W(C_{xx^\p}) W(C_{x^\p x})
\label{le}
\ee
where the coupling constant $\lambda=g^2N$ 
remains finite in the large-$N$ limit. 
The contours $C_{xx^\p}$ and $C_{x^\p x}$,
which are depicted in Fig.~\ref{Fig.1}, 
\begin{figure}[tbp]
\unitlength=1.00mm
\begin{picture}(44.0,43.00)(-40,90)
\thicklines
\bezier{60}(19.00,119.00)(27.00,120.00)(34.00,115.00)
\bezier{80}(11.50,108.00)(11.50,117.00)(19.00,119.00)
\bezier{80}(11.50,108.00)(11.50,100.00)(22.00,100.00)
\bezier{40}(22.00,100.00)(30.00,100.00)(33.00,103.00)
\bezier{64}(44.00,117.00)(48.00,119.00)(52.00,126.00)
\bezier{108}(52.00,126.00)(58.67,133.00)(60.67,117.50)
\bezier{148}(47.00,102.00)(62.67,103.00)(60.67,117.50)
\bezier{40}(44.00,117.00)(39.00,114.00)(37.00,110.00)
\bezier{64}(47.00,102.00)(40.00,103.00)(37.00,107.00)
\bezier{32}(34.00,115.00)(36.00,113.33)(37.00,110.00)
\bezier{32}(37.00,107.00)(34.67,104.33)(33.00,103.00)
\put(37.50,114.00){\makebox(0,0)[cb]{$x$}}
\put(37.00,104.00){\makebox(0,0)[ct]{$x^\prime$}}
\put(65.00,115.00){\makebox(0,0)[lc]{$C_{xx^\prime}$}}
\put(8.0,109.00){\makebox(0,0)[rc]{$C_{x^\prime x}$}}
\end{picture}
\caption[x]   {\hspace{0.2cm}\parbox[t]{13cm}
{\small
   Contours $C_{xx^\p}$ and $C_{x^\p x}$ which enter the r.h.s.\ 
   of Eqs.~\rf{le} and \rf{Le}. }}
\label{Fig.1}
\end{figure}
are the parts of the loop $C$ from 
$x$ to $x^\p$ and from $x^\p$ to $x$, respectively. They are always closed
due to the presence of the delta-function. It implies that 
$x$ and $x^\p$ should be the same points of {\it space}\/ but not necessarily
of the {\it contour} (\ie they may be associated with different values of the
parameter $\s$).  The principal value integral on the r.h.s.\ eliminates the 
case when $x$ and $x^\p$ are the same point of the contour. In other words the 
r.h.s.\ of \eq{le} does not vanish only if the contour $C$ has 
self-intersections.

To find $W(C)$, 
\eq{le} should be solved for the class of Stokes functionals with the initial 
condition 
\be
W(0)=1
\label{ic}
\ee
for loops which are shrunk to points.

\eq{le} is {\it not}\/ formulated, however, entirely on loop space. It 
is a $d$-vector equation whose both sides depend explicitly on the point $x$ 
which does not belong to loop space. The fact that one has a $d$-vector 
equation for the scalar quantity means, in particular, that \eq{le} is 
overspecified. 

A practical difficulty in solving \eq{le} is that area and path derivatives,
$ {\delta}/{\delta\s_{\mu\nu}( x)}$ and $\d_\mu^x$, respectively,
which enter the l.h.s.\ are complicated, generally speaking, noncommutive
operators. They are closely connected to the Yang--Mills perturbation theory 
where they correspond to the non-Abelian field strength~\rf{fieldstrength}
and the covariant derivative
\be
\nabla^x_\mu = \frac{\d}{\d x_\mu} + [A_\nu(x),~~~~]~.
\ee
It is not easy, however, to apply these operators to a generic functional 
$W(C)$ which is defined for the elements of loop space.

A much more convenient form of the loop equation can be obtained by integrating 
both sides of \eq{le} over $dx_\nu$ along the same contour, $C$, which yields
\be
\int_C dx_\nu \d_\mu^x \frac{\delta}{\delta\s_{\mu\nu}( x)} W(C) =
\lambda \int_C dx_\mu \pintc d x_\mu^\p \delta^{(d)}(x-x^\p) 
W(C_{xx^\p}) W(C_{x^\p x})\,.
\label{ile}
\ee
Now both the operator on the l.h.s.\ and the functional on the r.h.s.\ are 
scalars without labeled points 
and are well-defined on loop space. The operator on the l.h.s.\ 
of \eq{ile} can be interpreted as an infinitesimal variation of the elements 
of loop space.

The proof of equivalence of scalar \eq{ile} and the original $d$-vector 
\eq{le} is based on the important property of \eq{le} both sides of which are 
annihilated identically by the operator $\d_\nu^x$~\cite{MM81}. Due to this 
property the vanishing of the contour integral of some vector is equivalent to 
vanishing of the vector itself so that \eq{le} can in turn 
be derived from \eq{ile}.
Subsect.~\ref{stochastic} contains an alternative proof of the equivalence 
of Eqs.~\rf{le} and \rf{ile}. 

\subsection{The loop-space Laplacian}

When applied to regular functionals which do not have marked points, 
the operator on the l.h.s.\ of \eq{ile} can be represented using 
Eqs.~\rf{vpath} and \rf{varea} in an equivalent form
\be
\Delta \equiv \int_{\s_0}^{\s_f} d\s  \int_{\s-0}^{\s+0} d\s^\p 
\frac{\delta}{\delta x_\mu(\s^\p)} \frac{\delta}{\delta x_\mu(\s)} ~.
\label{defDelta}
\ee
Such an operator was considered by Gervais and Neveu~\cite{GN79} who
first pointed out that the (classical) loop equation is nothing but a 
(homogeneous) functional Laplace equation~\cite{Lev51}
(see also Ref.~\cite{Fel86} for a recent review). 
\eq{ile} can be 
represented in turn as an (inhomogeneous) functional Laplace equation
\be
\Delta W(C) 
= \lambda \int_C dx_\mu \pintc d x_\mu^\p \delta^{(d)}(x-x^\p) 
W(C_{xx^\p}) W(C_{x^\p x})\,.
\label{Le}
\ee

It is worth noting that the representation~\rf{defDelta} of the functional
Laplacian is defined for a wider class of functionals than Stokes ones.
The point is that the standard definition of the functional 
Laplacian~\cite{Lev51} uses solely the concept of the second variation 
of a functional $U[x]$, namely the term in the second variation
which is proportional to $(\delta x_\mu(\s))^2$:
\be
\delta^2 U[x] = \frac 12 \int_{\s_0}^{\s_f} d\s U^{\p\p}_{xx}[x]
(\delta x_\mu(\s))^2 + \ldots \,.
\ee
The functional Laplacian $\Delta$ is then defined by the formula
\be
\Delta U[x] = \int_{\s_0}^{\s_f} d\s U^{\p\p}_{xx}[x] \,.
\ee
Here $U[x]$ can be an arbitrary, not necessarily
reparametrization-invariant functional. 
To emphasize this obstacle, we use the notation $U[x]$ for generic functionals
which are defined on $L_2$ space in comparison to $U(C)$ for the ones
which are defined on the elements of loop space.
As will be shown below, it is easier to deal with the whole operator
$\Delta$ than with the original area and path derivatives separately.

The functional Laplacian possesses a number of remarkable properties.
While the finite-dimensional Laplacian is an operator of the second order, 
the functional Laplacian is that of the first order and satisfies the 
Leibnitz rule
\be
\Delta ( U[x] V[x] ) = \Delta ( U[x] ) V[x] + U[x] \Delta ( V[x] ) \,.
\label{Leibnitz}
\ee

In the next section we shall approximate the functional Laplacian by a
(second-order) partial differential operator in such a way to preserve
some of these properties.

\subsection{Relation to stochastic quantization \label{stochastic}}

The equivalence of the loop equations~\rf{le} and \rf{ile} 
(or \rf{Le}) can be alternatively
shown~\cite{HM89} using the method of stochastic quantization.
This approach to Yang--Mills theory is based on the stochastic Langevin
equation~\cite{PW81}
\footnote{~We did not add here a gauge-fixing term~\cite{Zwa81} because it 
does not affect averages of gauge-invariant functionals.}
\be
-\d_t A^a_\nu(x,t)+ \nabla_\mu F^a_{\mu\nu} (x,t) = g^2 \eta^a_\nu(x,t)
\label{Langevin}
\ee
where the averaging over the (Gaussian) white noise is defined by
\be
\LA \eta_\mu^{ij}(x,t_1)\eta_\nu^{kl}(x,t_2)  \RA_\eta
= 2 \delta^{(d)}(x-y)\delta(t_1-t_2 ) \delta_{\mu\nu}
\left( \delta^{il}\delta^{kj}-\fr 1N \delta^{ij}\delta^{kl}  \right)\,.
\ee
Here $\mu,\nu= 1,\ldots,d$ and $a=1,\ldots,N^2$--$1$ while $t$ is the Markov 
time.

As is well-known, \eq{Langevin} results at equilibrium in the Schwinger--Dyson
equation
\be
\LA \left( \nabla_\mu F^a_{\mu\nu} (x) - g^2
\frac{\delta}{\delta A_\nu ^a (x)} \right) G^a[A] \RA_A = 0 
\label{f-oSD}
\ee
with $G^a[A]$ being some functional of the gauge field, $A$.
This equation is loosely called the first-order Schwinger--Dyson equation since
it results from \eq{Langevin} which involves the first variation of the action
w.r.t.\ the field. The loop equation~\rf{le} is nothing but \eq{f-oSD} \sloppy
with  $G^a[A] =\Phi^a[A,C_{xx}]$ where $\Phi^a[A,C_{xx}]$ is given by 
\eq{Wlmarked}.

Analogously, a second-order stochastic equation can be derived from the
Langevin equation~\rf{Langevin} for a (colorless) composite operator $G[A]$ 
which results at equilibrium in the second-order Schwinger--Dyson equation
\be
\LA \left( \int d^d x \nabla_\mu F^a_{\mu\nu} (x)
\frac{\delta}{\delta A_\nu ^a (x)} - {\Delta}_F \right) G[A] \RA_A = 0 
\label{s-oSD}
\ee
with
\be 
\Delta_F = g^2 \int d^{d} x \frac{\delta}{\delta A_\nu ^a (x)} 
\frac{\delta}{\delta A_\nu ^a (x)} \,.
\label{fLaplacian}
\ee
For the case when $G[A]$ coincide with the Wilson loop~\rf{Wilsonloop},
\eq{s-oSD} was first derived by Marchesini~\cite{Mar81}.

It can be shown~\cite{HM89} that the loop equation~\rf{Le} emerges after
inserting $G[A]=\Phi[A,C]$ with $\Phi[A,C]$ given by \eq{Wl} into \eq{s-oSD}, 
calculating the variational derivatives and translating the result into 
loop space. Therefore, \eq{Le} is associated with the second-order 
Schwinger--Dyson equation exactly in the same sense as \eq{le} is associated 
with the first-order one.

It is interesting to note that the operator $\Delta_F$ on the l.h.s.\ of 
\eq{s-oSD} has the meaning of the functional Laplacian on the field space.
The loop equation~\rf{Le} can be rewritten, therefore, as the relation
\be
\LA \Big( \Delta + \Delta_F \Big) \Phi[A,C] \RA_A =0
\ee
between the Laplacians on loop and field spaces.

\subsection{Nonperturbative regularization}

A convenient way of ultraviolet regularization of the loop equation~\rf{Le} was 
proposed by Halpern and the author~\cite{HM89} using the relation to the 
stochastic quantization discussed in the previous subsection. 

The idea is to 
start from the regularized version of the Langevin equation~\cite{BHST87}
\be
-\d_t A^a_\nu(x,t)+ \nabla_\mu F^a_{\mu\nu} (x,t) = 
g^2 \int dy R(\nabla^2)^{ab}_{xy} \eta^b_\nu(y,t)
\label{rLangevin}
\ee
with the regularizing operator 
\be
R(\nabla^2)^{ab}_{xy} = \left( \e^{\frac{\nabla^2}{4\Lambda^2}}  \right)^{ab} 
\delta^{(d)}(x-y) \,.
\label{defR}
\ee
The regularized version of \eq{s-oSD} is
\be
\LA \left( \int d^d x \nabla_\mu F^a_{\mu\nu} (x)
\frac{\delta}{\delta A_\nu ^a (x)} - {\Delta}^R_F \right) G[A] \RA_A = 0 
\label{rs-oSD}
\ee
with
\be 
\Delta_F^R = g^2 \int d^{d} x d^d y (R^2(\nabla^2))^{ab}_{xy} 
\frac{\delta}{\delta A_\nu ^b (y)} 
\frac{\delta}{\delta A_\nu ^a (x)} \,.
\label{rfLaplacian}
\ee

To translate \eq{rs-oSD} into loop space, one uses the path-integral 
representation 
\be
(R^2)^{ab}_{xy} = \int_{r(0)=x}^{r(\Lambda^{-2})=y} {\cal D} r
\e^{-\fr 12 \int_0^{\Lambda^{-2}} d \tau \dot{r}^2(\tau)}
\tr{}\left[ t^a U(r_{xy})
t^b  U(r_{yx})
   \right] \label{rpi} 
\ee
with
\be
U(r_{xy}) =
\P \e^{\int_0^{\Lambda^{-2}} d \tau \dot{r}_\mu(\tau) A_\mu(r(\tau))}
\ee
where the integration is over regulator paths $r_\mu(\tau)$ from $x$ to $y$
whose typical length is $\sim \Lambda^{-1}$. 
The conventional measure is implied in~\rf{rpi} so that
\be
 \int_{r(0)=x}^{r(\Lambda^{-2})=y} {\cal D} r
\e^{-\fr 12 \int_0^{\Lambda^{-2}} d \tau \dot{r}^2(\tau)}
\tr{}\left[ t^a t^b  \right] 
=\delta ^{ab} \e^{\frac{\Box}{2\Lambda^2}} \delta^{(d)}(x-y)
=\delta^{ab} \left(\frac{\L^2}{2\pi}\right)^{\frac d2}
\e^{-\frac{(x-y)^2 \L^2}{2}}
\label{rpi0} 
\ee
where $\Box$ stands for the $d$-dimensional Laplace operator.

As was shown in Ref.~\cite{HM89}, the application of the regularized 
operator~\rf{rfLaplacian} to the Wilson loop yields as $N\ra\infty$:
\be
- \Delta^R_F \Phi[A,C] = \l \oint_C dx_\mu \oint_C dy_\mu
 \int_{r(0)=x}^{r(\Lambda^{-2})=y} {\cal D} r
\e^{-\fr 12 \int_0^{\Lambda^{-2}} d \tau \dot{r}^2(\tau)}
\Phi[A,C_{xy}r_{yx}]\Phi[A,C_{yx}r_{xy}] \,,
\label{AnLe}
\ee
where the contours $C_{xy}r_{yx}$ and $C_{yx}r_{xy}$ are depicted in
Fig.~\ref{Fig.2}.
\begin{figure}[tbp]
\unitlength=1.00mm
\begin{picture}(44.0,43.00)(-40,90)
\thicklines
\bezier{60}(20.00,119.00)(28.00,120.00)(35.00,115.00)
\bezier{80}(12.50,108.00)(12.50,117.00)(20.00,119.00)
\bezier{80}(12.50,108.00)(12.50,100.00)(23.00,100.00)
\bezier{40}(23.00,100.00)(31.00,100.00)(34.00,103.00)
\bezier{64}(44.00,117.00)(48.00,119.00)(52.00,126.00)
\bezier{108}(52.00,126.00)(58.67,133.00)(60.67,117.50)
\bezier{148}(47.00,102.00)(62.67,103.00)(60.67,117.50)
\bezier{40}(44.00,117.00)(39.00,114.00)(38.00,111.00)
\bezier{64}(47.00,102.00)(40.00,103.00)(38.00,106.00)
\bezier{32}(35.00,115.00)(37.00,113.33)(37.00,111.00)
\bezier{32}(37.00,106.00)(35.67,104.33)(34.00,103.00)
\put(38.0,114.00){\makebox(0,0)[cb]{$x$}}
\put(38.0,104.00){\makebox(0,0)[ct]{$y$}}
\put(65.00,115.00){\makebox(0,0)[lc]{$C_{xy}$}}
\put(9.0,109.00){\makebox(0,0)[rc]{$C_{y x}$}}
\thinlines
\bezier{28}(38.00,111.00)(36.00,109.00)(38.00,108.00)
\bezier{28}(38.00,108.00)(39.00,107.00)(38.00,106.00)
\bezier{28}(37.00,111.00)(35.00,109.00)(37.00,108.00)
\bezier{28}(37.00,108.00)(38.00,107.00)(37.00,106.00)
\put(34.00,109.00){\makebox(0,0)[rc]{$r_{xy}$}}
\put(40.00,109.00){\makebox(0,0)[lc]{$r_{yx}$}}
\end{picture}
\caption[x]   {\hspace{0.2cm}\parbox[t]{13cm}
{\small
   Contours $C_{xy}r_{yx}$ and $C_{y x}r_{xy}$ which enter the r.h.s.\ 
   of Eqs.~\rf{AnLe} and \rf{rLe}. }}
\label{Fig.2}
\end{figure}
Averaging over the gauge field, $A$, one arrives at the regularized loop 
equation
\be
\Delta W(C) =  \l \oint_C dx_\mu \oint_C dy_\mu
 \int_{r(0)=x}^{r(\Lambda^{-2})=y} {\cal D} r
\e^{-\fr 12 \int_0^{\Lambda^{-2}} d \tau \dot{r}^2(\tau)}
W(C_{xy}r_{yx}) W(C_{yx}r_{xy}) 
\label{rLe}
\ee
which manifestly recovers \eq{Le} when $\L\ra\infty$.

The constructed regularization is nonperturbative while perturbatively 
should recover the regularization of Feynman diagrams of Ref.~\cite{BHST87} 
which is based on smearing the noise in the
method of stochastic quantization. The 
advantage of this regularization of the loop equation is that the contours
$C_{xy}r_{yx}$ and $C_{y x}r_{xy}$ on the r.h.s.\ of \eq{rLe} both are closed
and do not have marked points if $C$ does not have.
Therefore, \eq{rLe} is written entirely on loop space.

\subsection{Extensions to scalars, ${\cal N}=4$ SYM and all that}

As is already mentioned, the functional Laplacian is defined for a wider class of
 functionals than the functionals of Stokes' type. It can be thus applied to a scalar 
 Wilson loop
 \be
 U[\phi,x]=\frac{{\rm tr}}N \Big(\hbox{P}\e^{\mu \int d \sigma \sqrt{\dot x^2(\sigma)} \phi (x(\sigma))}\Big)
 \label{Pphi}
 \ee
 where $\phi$ is generically a matrix-valued scalar field and $\mu$ 
 is a constant of dimension of  $[\hbox{mass}]^{2-d/2}$. 
The path-ordered exponential in 
 \rf{Pphi}  is parametric invariant but does not obey the backtracking condition as Stokes' functionals do.
  
The case of one free scalar was elaborated in Ref.~\cite{Mak88}. 
Using the definition~\rf{defDelta}, it is straightforward to show that the average 
\be
W[x]=\LA  U[\phi,x] \RA _\phi
\ee
 over $\phi$ 
with the free action of \rf{Pphi} satisfies the simple loop equation
\be
\Delta W[x]
= -\mu^2 \int d \sigma \pint d \sigma^\p \sqrt{\dot x^2(\sigma)} \sqrt{\dot x^2(\sigma^\p)} 
\delta^{(d)}\big(x(\sigma)-x(\sigma^\p)\big)
W[x]
\label{sLe}
\ee
whose solution is
\be
W[x]=\exp\Big({\frac{\mu^2}2  \int d \sigma \pint d \sigma^\p \sqrt{\dot x^2(\sigma)} \sqrt{\dot x^2(\sigma^\p)} D_0\big(x(\sigma)-x(\sigma^\p)\big)} \Big)
\ee
whith $D_0$ being the free scalar propagator.

The Wilson loops~\rf{Pphi} and \rf{Wl} are constituents of the Wilson loops in ${\cal N}=4$ SYM
\cite{Mal98b,RY98}. They obey a supersymmetric extension of the loop equation~\cite{DGO99}
which deals with supersymmetric loops
${\bbox C}=\{x_\mu(\sigma), Y_i(\sigma);
\zeta(\sigma)\}$, where $\zeta(\sigma)$ denotes the Grassmann odd component
and the supersymmetric functional Laplacian reads
\be {\bbox \Delta} = \int d \s
\int_{\s-0}^{\s+0} d \s' \left( \frac{\delta^2}{\delta
x^\mu(\s')\delta x_\mu(\s)}+ \frac{\delta^2}{\delta Y^i(\s')\delta
Y_i(\s)}+ \frac{\delta^2}{\delta \zeta(\s')\delta \bar\zeta(\s)}\right).
\label{2.41}
\ee

An interesting application of 
 the supersymmetric loop equation is for the cusp anomalous dimension at the 
 one-loop~\cite{DGO99}
 and two-loop~\cite{MOS06} orders of perturbation theory
 (sorry for the pun). 
 Then we can set $\dot Y^2=\dot x^2$, $\zeta=0$ after acting by
${\bbox \Delta} $ to have
\bea {\bbox
\Delta} W({\bbox C})\Big|_{{\bbox C}=\Gamma} &= &\lambda \int d
\sigma \pint d \sigma^\p \, \left( \dot x_\mu(\sigma)\dot
x_\mu(\sigma^\p)- \sqrt{\dot x^2(\sigma)} \sqrt{\dot x^2(\sigma^\p)} 
\right) \non &&~~~~~\times
\delta^{(4)}{\big(x(\sigma)-x(\sigma')\big)}{W(\Gamma_{x(\sigma)
x(\sigma')})W(\Gamma_{x(\sigma')x(\sigma)})} \,, \label{cle} \eea
where $\Gamma$ denotes the loop with a cusp.
The coefficient on the right-hand side of
\eq{cle} accounts for the fact that the Wilson loops are in the
adjoint representation.  A different
but equivalent operator was used in Ref.~\cite{FKKT97,IIB} for deriving
loop equations in the IIB matrix model as is briefly discussed in Subsect.~\ref{s:regul} below.



\newsection{The polygon discretization\label{s2}}

To treat the functional Laplace equation, it is useful to discretize 
loop space  
by $M$-vertex polygons and approximate the loop-space Laplacian
by a (finite-dimensional) second-order partial
differential operator of a specific form. 
The loop equation can then be approximated by a
by a second-order partial differential equation. 
Among other issues, the polygon
discretization allows to obtain the Green function of the 
approximating operator which is
represented in the form of a Gaussian integral.
Therefore, the
loop-space Laplace equation can be transformed to an equivalent
integral form. The continuum limit of the discretized Green function exists as 
$M\ra\infty$ and takes the form of a path integral for the Euclidean
harmonic oscillator.

\subsection{Discretization of loop space \label{ss3.1}}

\begin{figure}[tbp]
\unitlength=1.00mm
\begin{picture}(44.0,43.00)(-50,90)
\thicklines
\bezier{80}(11.5,108.00)(11.5,119.00)(19.00,119.00)
\bezier{64}(19.00,119.00)(28.00,121.00)(32.00,126.00)
\bezier{108}(32.00,126.00)(38.67,133.00)(40.67,117.50)
\bezier{148}(21.00,100.00)(42.67,101.00)(40.67,117.50)
\bezier{80}(11.5,108.00)(11.5,99.50)(22.00,100.00)
\thinlines
\put(31.00,125.00){\line(4,1){7.33}}
\put(38.33,126.67){\line(1,-4){2.57}}
\put(41.00,116.33){\line(-1,-2){6.67}}
\put(34.33,103.00){\line(-4,-1){12.33}}
\put(22.00,100.00){\line(-2,1){10.00}}
\put(31.00,125.00){\line(-2,-1){12.00}}
\put(19.00,119.0){\line(-1,-2){7.0}}
\put(22.00,97.00){\makebox(0,0)[ct]{$x_1$}}
\put(36.00,101.00){\makebox(0,0)[lt]{$x_M$}}
\put(17.00,121.00){\makebox(0,0)[rb]{$x_{i-1}$}}
\put(30.00,127.00){\makebox(0,0)[rb]{$x_i$}}
\put(40.00,128.00){\makebox(0,0)[lb]{$x_{i+1}$}}
\end{picture}
\caption[x]   {\hspace{0.2cm}\parbox[t]{13cm}
{\small
   Approximation of a continuous loop by the $M$-gon. 
    }}
\label{Fig.3}
\end{figure}
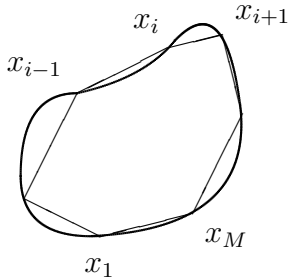

Loop space (which is a functional one) 
can be discretized by the space of polygons with 
$M$ vertices. Any continuous loop can be approximated by the $M\ra\infty$ limit 
of a $M$-vertex polygon.
This is done by choosing arbitrary $M$ values $\s_i\in[\s_0,\s_f]$
($i=1,\ldots,M$) so that $\s_{i-1}<\s_{i}$, $\s_0=\s_M$
and identifying the $i$-th vertex $x_i$ of the $M$-gon with 
the point $x(\s_i)$ lying on the loop (see Fig.~\ref{Fig.3}).  
It is evident that any
continuous loop can be approximated by such $M$-gons providing the distance 
$\delta s_i=\sqrt{(x_i-x_{i-1})^2}$ ($x_0=x_M$)
between neighbor points vanishes as $M\ra\infty$. The smooth loops 
are characterized by $\delta s_i = {\cal O}(M^{-1})$.

The functionals $W(C)$ which are defined on the elements of loop space (and 
can be denoted alternatively as $W[x]$) can be approximated by the 
functions $W(x_1,\ldots,x_M)$ of $d\cdot M$ variables --- the vertices of the 
$M$-gon --- providing these functions obey several relations which 
guarantee the desired functional limit.
Some of them --- the symmetry w.r.t.\ to a cyclic permutation of the variables 
or their simultaneous shift by a constant vector --- can easily be formulated 
at finite $M$. Other relations, which are associated with the fact that 
the functional $W[x]$ does not have marked points or is invariant under 
reparametrizations, are formulated as $M\ra\infty$:
\vspace{-10pt}
\begin{itemize}
\addtolength{\itemsep}{-8pt}
\item[1.] 
If for $|x_i-x_{i-1}|$
\be
\frac{\d}{\d x_i} W (x_1,\ldots,x_M) = {\cal O}\left( M^{-1} \right)
~~~~~~~(i=1,\ldots,M)   ~,
\ee
$W (x_1,\ldots,x_M)$ recovers a functional without marked points as
$M\ra\infty$. 

\item[2.] 
If simultaneously, say,
\be
(x_i-x_{i-1})_\mu \frac{\d}{\d x_i^\mu} W (x_1,\ldots,x_M) =
 {\cal O}\left( M^{-3} \right) ~~~~~~~(i=1,\ldots,M)   ~, 
\ee
the limiting functional will be reparametrization-invariant.

\item[3.]
If $W (x_1,\ldots,x_M)$ obeys in addition a discrete analogue of the 
backtracking condition~\rf{btr}, the limiting functional will be the one of the 
Stokes type.

\end{itemize}
\vspace{-9pt}

The explicit form of these relations will not be essential for what follows.
The point is that $W(x_1,\ldots,x_M)$ which will be calculated as 
a solution of 
the discretized version of the functional Laplace equation~\rf{Le} (see 
Eqs.~\rf{3.13}, \rf{3.12}  below) automatically obeys these constraints 
providing the discretization of the functional on the r.h.s.\ of \eq{Le} does. 
The explicit discretized form of it (given by \eq{3.13} below) possesses this 
property. 

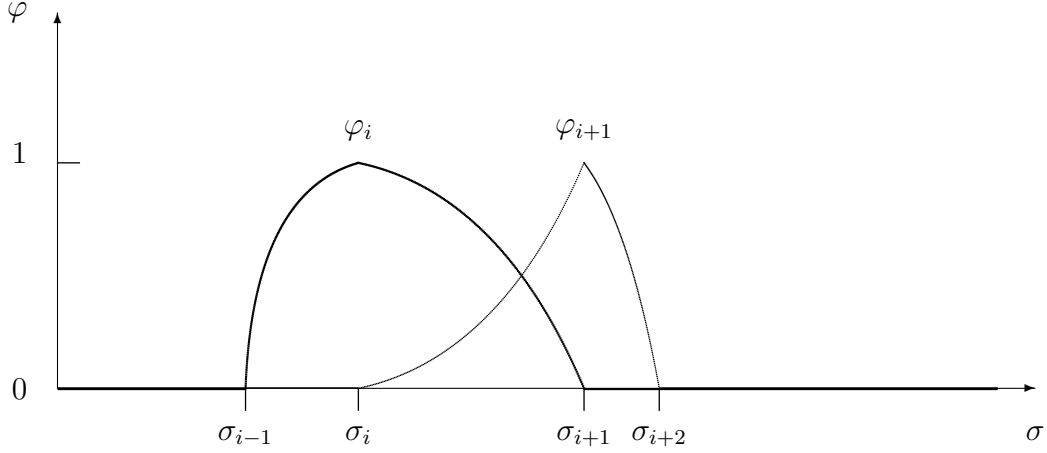
\begin{figure}[tbp]
\unitlength=1.00mm
\begin{picture}(44.0,63.00)(-10,70)
\thinlines
\put(10.00,80.00){\vector(1,0){130.00}}
\put(10.00,80.00){\vector(0,1){50.00}}
\thicklines
\put(10.00,79.9){\line(1,0){25.00}}
\bezier{168}(35.00,80.00)(36.00,106.00)(50.00,110.00)
\bezier{184}(50.00,110.00)(69.00,106.00)(80.00,80.00)
\put(80.00,79.9){\line(1,0){55.00}}
\linethickness{0.5pt}
\bezier{184}(80.00,110.00)(69.00,84.00)(50.00,80.00)
\bezier{168}(80.00,110.00)(86.00,102.00)(90.00,80.00)
\put(10.00,80.05){\line(1,0){40.00}}
\put(90.00,80.05){\line(1,0){45.00}}
\put(10.00,110.00){\line(1,0){3.00}}
\put(35.00,80.00){\line(0,-1){3.00}}
\put(80.00,80.00){\line(0,-1){3.00}}
\put(50.00,80.00){\line(0,-1){3.00}}
\put(90.00,80.00){\line(0,-1){3.00}}
\put(50.00,113.00){\makebox(0,0)[cb]{$\varphi_i$}}
\put(6.00,130.00){\makebox(0,0)[rc]{$\varphi$}}
\put(6.00,110.00){\makebox(0,0)[rb]{$1$}}
\put(6.00,80.00){\makebox(0,0)[rc]{$0$}}
\put(35.00,75.00){\makebox(0,0)[ct]{$\sigma_{i-1}$}}
\put(50.00,75.00){\makebox(0,0)[ct]{$\sigma_i$}}
\put(80.00,75.00){\makebox(0,0)[ct]{$\sigma_{i+1}$}}
\put(90.00,75.00){\makebox(0,0)[ct]{$\sigma_{i+2}$}}
\put(140.00,75.00){\makebox(0,0)[ct]{$\sigma$}}
\put(80.00,113.00){\makebox(0,0)[cb]{$\varphi_{i+1}$}}
\end{picture}
\caption[x]   {\hspace{0.2cm}\parbox[t]{13cm}
{\small
   Basis functions $\varphi_i(\s)$ which enter \eq{basis}. 
    }}
\label{Fig.4}
\end{figure}

One might ask the inverse question: how the continuous loop $x(\s)$ can be 
constructed for a given set of $M$ points $\{x_1,\ldots,x_M \}$?
Let us split again the interval $\s_0,\s_f$ into $M$ 
(maybe nonequal) intervals
$[\s_{i-1},\s_i]$ and identify the point $x_i$ with the $i$-th vertex of a 
$M$-gon. Let us define the function $x_\mu(\s)$, which runs along the $M$-gon
when the parameter $\s$ is increased from $\s_o$ to $\s_f$, by the formula
\be
x(\s) = \sum _{i=1}^M x_i \varphi_i(\s) \,.
\label{basis}
\ee
The function $\varphi_i(\s)$ is depicted in Fig.~\ref{Fig.4}.
It is nonvanishing only at the $i$-th and ($i$+$1$)-th intervals while
\be
\varphi_{i+1}(\s) = 1 - \varphi_i(\s)~~~~~~~\hbox{for}~~~~\s\in[\s_i,\s_{i+1}]
\ee
and $\varphi^\p_{i+1}(\s)\geq 1$ at the $i$-th interval. This function is 
arbitrary otherwise. 
The independence of results on the form of $\varphi_i(\s)$ as well as on the 
choice of $\s_i$ is a remembrance of the reparametrization invariance at finite 
$M$.

There is an important difference between the polygon discretization and the 
standard approximation~\cite{Lev51}
of $L_2$-space by the stepwise functions: $x(\s)=x_i$ 
for $\s\in[\s_{i-1},\s_i]$, $x(\s)=0$ otherwise. For the polygon 
discretization the approximating functions are continuous at any $M$ while for 
the stepwise one they become continuous only when $M=\infty$. Since in a gauge 
theory the continuity of loops is associated with gauge invariance, we see an 
advantage of the polygon discretization.

Thus, the formula~\rf{basis} has at $M\ra\infty$ the meaning of the expansion 
over the whole set of basis functions $\varphi_i(\s)$. In contrast to the 
standard expansion over the stepwise functions~\cite{Lev51}, our basis 
functions are not orthogonal at finite $M$. Nevertheless, their linear 
combinations can be made orthogonal.

\subsection{Discretized functional Laplacian}

The finite-dimensional discretization of the functional Laplacian 
on loop space is given by the following second-order operator
\be
\Delta^{(M)} = \sum_{i,j=1}^M \frac{\d}{\d x_i^\mu} Q_{ij}
\frac{\d}{\d x_j^\mu}
\label{discDelta}
\ee
which involves the partial derivatives w.r.t.\ the coordinates of  
the vertices of the $M$-gon. 
Due to the cyclic symmetry, $Q_{ij}$ is an $M\times M$
Toeplitz matrix which depends only on
$|i-j|$ defined \ mod$\,M$:
\be
Q_{i,i\pm K} = Q_K ~~(K=0,1,\ldots, M-1)\,, ~~~~~Q_K=Q_{M-K}
\label{3.2}
\ee
or explicitly
\be
Q_{ij}=\left\{
        \begin{array}{ccccccccccccc}
                        Q_0&Q_1&Q_2&Q_3&Q_4&Q_5&.&.&.&Q_4&Q_3&Q_2&Q_1 \\*
                        Q_1&Q_0&Q_1&Q_2&Q_3&Q_4&.&.&.&Q_5&Q_4&Q_3&Q_2 \\*
                        Q_2&Q_1&Q_0&Q_1&Q_2&Q_3&.&.&.&Q_6&Q_5&Q_4&Q_3 \\*
                        Q_3&Q_2&Q_1&Q_0&Q_1&Q_2&.&.&.&Q_7&Q_6&Q_5&Q_4 \\*
                        Q_4&Q_3&Q_2&Q_1&Q_0&Q_1&.&.&.&Q_8&Q_7&Q_6&Q_5 \\*
                        Q_5&Q_4&Q_3&Q_2&Q_1&Q_0&.&.&.&Q_9&Q_8&Q_7&Q_6 \\*
                        .  &.  &.  &.  &.  &.  &.&.&.&.  &.  &.  &.   \\*
                        .  &.  &.  &.  &.  &.  &.&.&.&.  &.  &.  &.   \\*
                        .  &.  &.  &.  &.  &.  &.&.&.&.  &.  &.  &.   \\*
                        Q_4&Q_5&Q_6&Q_7&Q_8&Q_9&.&.&.&Q_0&Q_1&Q_2&Q_3 \\*
                        Q_3&Q_4&Q_5&Q_6&Q_7&Q_8&.&.&.&Q_1&Q_0&Q_1&Q_2 \\*
                        Q_2&Q_3&Q_4&Q_5&Q_6&Q_7&.&.&.&Q_2&Q_1&Q_0&Q_1 \\*
                        Q_1&Q_2&Q_3&Q_4&Q_5&Q_6&.&.&.&Q_3&Q_2&Q_1&Q_0
        \end{array}  \right\}\;.
\label{toeplits}
\ee

The simplest (but not unique) choice of $Q_K$ is~\cite{Mak88} 
\be
Q_K = \frac{q^K+q^{M-K}}{1-q^M} \,.
\label{Qq}
\ee
It guarantees, as is shown below, that $\Delta^{(M)}$ tends to the functional 
Laplacian as $M\ra\infty$ if the number $q<1$ satisfies
\be
q\ra1~,~~~~~~~M(1-q)\ra\infty~~~~~~~~\hbox{as}~~M\ra\infty\,.
\label{3.3}
\ee
The term $q^{M-K}$ in the numerator of~\rf{Qq} guarantees $Q_K=Q_{M-K}$ while
the term $q^M$ in the denominator simplifies the inverse matrix 
$Q_{ij}^{-1}$ (given by \eq{4.3} below).

An explicit form of the discretization~\rf{discDelta}, \rf{Qq} for large $M$ 
reads 
\be 
\Delta^{(M)} = \sum_{i=1}^M \left( \frac{\d}{\d x_i} \frac{\d}{\d x_i} 
+ 2 q \frac{\d}{\d x_i} \frac{\d}{\d x_{i-1}} + \ldots 
+ 2 q^K \frac{\d}{\d x_i} \frac{\d}{\d x_{i-K}} + \ldots \right) 
\label{3.4} 
\ee 
where 
the sum over $K$ contains $[M/2]+1$ terms.  However, only $K\sim(1-q)^{-1}$ 
terms are non-vanishing in the limit~\rf{3.3}.  In this sense our 
discretization of the functional Laplacian is local while the r.h.s.\ of 
\eq{3.4} contains not only partial derivatives w.r.t.\ nearest neighbor 
vertices. This locality were absent if $M(1-q)\sim 1$ rather than 
$M(1-q)\ra \infty$.

It is worth mentioning that $\Delta^{(M)}$, which is  a second-order 
differential operator at finite $M$, becomes that of first order (satisfies the 
Leibnitz rule~\rf{Leibnitz}) as $M\ra\infty$ when applied to functionals 
without marked points.

\subsubsection{Why such $Q_{ij}$?}

There are several explanations why the discretization~\rf{discDelta} of the 
functional Laplacian with $Q_{ij}$ given by Eqs.~\rf{3.2}, \rf{Qq}
is adequate. Let us demonstrate that the second-order 
partial differential operator $\Delta^{(M)}$ tends to the functional Laplacian 
$\Delta$ as $M\ra\infty$ and $q\ra1$ according to~\rf{3.3}.

To this aim, let us apply the operator~\rf{discDelta} to the discretized Wilson 
loop
\be
\Phi(x_1,\ldots,x_M) = \ntr \prod_{i=1}^{M} U_i (x_{i-1},x_i)
\label{discPhi}
\ee
where
\be
 U_i(x_{i-1},x_i) = \P_{\tau} \e^{\int_0^1 d \tau \mu_i(\tau) A_\mu(x(\tau))
 (x_i-x_{i-1})_\mu}
\label{Ui}
\ee
and $x(\tau)=x_{i-1}+\tau (x_i-x_{i-1})$ runs over the $i$-th link of the 
$M$-gon when $\tau \in [0,1]$.

If all $\mu_i(\tau)=1$, \rf{discPhi} coincides with the usual phase factor for
the contour being the $M$-gon.
For an arbitrary $\mu_i(\tau)$ \rf{discPhi} differs from the phase factor but
tends to it when $\delta s_i=\sqrt{(x_i-x_{i-1})^2}\ra0$ as $M\ra\infty$
providing
\be
\int_0^1 d \tau \mu_i(\tau) =1 \,.
\label{3.7}
\ee
To verify the proposed formulas, we shall consider the case of arbitrary 
weights $\mu_i(\tau)$ which obey~\rf{3.7} as well as the condition \be 
\mu_i(\tau)=\mu_i(1-\tau)  
\label{3.8}
\ee
which guarantees the Stokes property
\be
 U_i(x_{i-1},x_i) U_i(x_i,x_{i-1}) =1
\ee
even at finite $M$. Our criterion of correct formulas will be an independence
of results on the choice of $\mu_i(\tau)$ as $M\ra\infty$.

Since we are interested in the limiting case $\delta s_i\ra 0$ as $M\ra0$, the
r.h.s.\ of \eq{Ui} can be expanded in $\delta s_i$:
\be
U_i(x_{i-1},x_i) = 1 +A_\cdot + \fr 12 A_\cdot^2 + \fr 16 A_\cdot^3 +
\fr {c_1}{24}\d_\cdot^2 A_\cdot 
-\fr{c_2}{12} [\d_\cdot A_\cdot,\,A_\cdot] + {\cal O}(\delta s_i^4) \,,
\label{3.10}
\ee
where the vector potential is taken at the point $(x_i+x_{i-1})/{2}$ 
--- the middle of the $i$-th link of the $M$-gon:
\be
A_\cdot \equiv (x_i-x_{i-1})_\mu A_\mu\left( \fr{x_i+x_{i-1}}{2} \right)~,
\ee
and the derivative
\be
\d_\cdot \equiv (x_i-x_{i-1})_\mu \d_\mu 
\ee
is w.r.t.\ this point. The constants $c_1$ and $c_2$ are expressed via some
integrals of the weight function $\mu_i(\tau)$ while $c_1=c_2=1$ for
$\mu_i(\tau)=1$. Therefore, our criterion of correct formulas will 
transform into an independence of results on $c_1$ and $c_2$.

Let us now apply the operator~\rf{discDelta} to the ansatz~\rf{discPhi},
\rf{3.10}. The result involves two terms: the first one with the double sum 
over the links of the  $M$-gon and the second term with the single sum
which involves four terms in the parenthesises.
The result of differentiation simplifies for a smooth contour when all $\delta 
s_i \sim 1/M$ as $M\ra\infty$. One gets (for an arbitrary $Q_{ij}$) 
\bea 
\Delta^{(M)} \Phi\! &=& \!
\sum_i \sum_j Q_{|i-j|}(x_i-x_{i-1})_\mu (x_j-x_{j-1})_\nu
\ntr{} \left\{ \P \prod_{l=1}^M U_l 
F_{\l\mu}\left( \fr{x_i+x_{i-1}}{2} \right) 
F_{\l\nu}\left( \fr{x_j+x_{j-1}}{2} \right) \!\right\} \non
 & &+ \sum_i (x_i-x_{i-1})_\nu
\ntr{} \Big\{ \P \prod_{l=1}^M U_l 
\left[ \fr{(2Q_0+Q_1)}{3}\nabla_\mu F_{\mu\nu}\left(\fr{x_i+x_{i-1}}{2} \right) 
+  \fr{(c_1-1)(Q_0-Q_1)}{6} \d^2 A_\nu 
\right. \non & & \left.
+ \fr{(c_1-c_2)(Q_0-Q_1)}{3} [\d_\mu A_\mu,\,A_\nu] 
+\fr{(c_2-1)(Q_1-Q_0)}{3} [(\d_\nu A_\mu+\d_\mu A_\nu),\, A_\mu ] 
\right] \Big\}\,.
\label{3.11}
\eea
Notice that the last three terms in the parenthesises, 
which are not gauge-invariant, vanish for
$c_1=c_2=1$ when~\rf{discPhi} coincides with the gauge-invariant phase factor.

Let us estimate the order in $1/M$ of each term on the r.h.s.\ of \eq{3.11}
for the choice~\rf{Qq} of $Q_K$ in the limit~\rf{3.3}.
The first term (with the double sum) is of order $(M(1-q))^{-1}$ 
for smooth contours and vanishes in the limit~\rf{3.3} in full analogy with the 
approximations of functional Laplacians of Ref.~\cite{Lev51}. This were not be 
the case for $M(1-q)\sim1$. The term with the single sum over 
$i$ recovers the functional Laplacian with the coefficient $(2+q)/3\ra1$ plus 
three more terms which vanish as $q\ra1$.

These last three terms would not vanish for arbitrary $c_1$ and $c_2$ if $q$
were not tend to $1$.  This occurs, for instance, for the simplest
discretization of the functional Laplacian by ordinary Laplacian in $d\cdot
M$-dimensional Euclidean space which is associated with $q=0$. For such $q$
the terms of order ${\cal O}(M^{-3})$ in the expansion~\rf{3.10} would 
``revive'' under the action of $\Delta^{(M)}$ and give nonvanishing 
contribution. Therefore, our criterion for the results to be independent 
of $c_1$ and $c_2$ fixes unambiguously that $q\ra1$ as $M\ra\infty$ while
vanishing of the first term (with the double sum) requires that simultaneously 
$M(1-q)\ra\infty$.

One more explanation of ``why such $Q_{ij}$''  is given in Subsect.~\ref{ss3.4}

\subsection{Discretized loop equation}

We use the above discretization~\rf{discDelta}, \rf{Qq}, \rf{3.3} of 
the loop-space Laplacian to discretize the l.h.s.\ of the loop 
equation~\rf{Le}. The r.h.s.\ can be approximated by  
\bea
J(x_1,\ldots,x_M)\!&= &\!\l \sum_i \sum_j (x_i-x_{i-1})_\mu (x_j-x_{j-1})_\mu \,
\delta^{(d)} \left( \fr{x_i+x_{i-1}}{2} - \fr{x_j+x_{j-1}}{2} \right) \non&&~~~\times
W(x_i,x_{i+1}, \ldots, x_{j-2},x_{j-1})\,  W(x_j,x_{j+1}, \ldots, 
x_{i-2},x_{i-1}) ~~~~
\label{3.13} 
\eea
so that the discretized loop equation reads
\be 
\Delta^{(M)} W(x_1,\ldots, x_M) = J(x_1,\ldots,x_M) \,.
\label{3.12}
\ee
                                                           
Some comments concerning the discretization~\rf{3.13} are in order:
\vspace{-10pt}
\begin{itemize}
\addtolength{\itemsep}{-8pt}
\item[i)] 
\rf{3.13} recovers the r.h.s.\ of \eq{Le} as $M\ra\infty$ preserving at 
finite $M$ some properties which are mentioned in Subsect.~\ref{ss3.1}.

\item[ii)] 
Other approximations of the r.h.s.\ of \eq{Le} can be done with the differences
vanishing as $M\ra\infty$ for the smooth contours. 

\item[iii)]
A convenient approximation is when the region $|i-j|\la (1-q)^{-1}$ is 
eliminated from the sum over $j$ on the r.h.s.\ of \eq{3.13}. This corresponds 
to the principal value integral over $dx^\p$ on the r.h.s.\ of \eq{Le}.

\end{itemize}
\vspace{-9pt}

Some properties of the discretized loop equation~\rf{3.12} are discussed in the 
next subsection where it is transformed to an integral form.

\subsection{Discretized Green function \label{ss3.4}}

Let us consider the current $J(x_1,\ldots,x_M)$ on the r.h.s.\ of \eq{3.12} to 
be known. This takes place, say, for the Abelian loop equation or when one 
solves the non-Abelian one by iterations in the coupling constant, $\l$. The 
solution can conveniently be written down using the Green function of the 
operator $\Delta^{(M)}$.

The operator $\Delta^{(M)}$ can easily be inverted with the boundary
condition~\rf{ic} to give
\bea
W(x_1,\ldots,x_M) = 1 - \half \int_0^\infty dA \left\{
\LA J(x_1+\sqrt{A}\xi_1,\ldots, x_M+\sqrt{A}\xi_M ) \RA _\xi^{(M)} \right.\non
- \left. \LA J(\sqrt{A}\xi_1,\ldots, \sqrt{A}\xi_M ) \RA _\xi^{(M)}
\right\}\hspace*{1.2cm}
\label{4.1}
\eea
where the parameter $A$ of the dimensionality of $[\,$length$\,]^2$ has the 
meaning of the proper time during which the evolution occurs from the initial 
$M$-gon with the coordinates $x_i$ to the final one with the coordinates 
$x_i+\sqrt{A} \xi_i$.
The averaging over the variables $\xi_i$ ($i=1,\ldots,M$) is Gaussian:
\be
\BLA F(\xi_1,\ldots, \xi_M ) \BRA _\xi^{(M)} \equiv
\frac{\int \prod_{i=1}^M d^d \xi_i
\e^{-\frac 12 \sum_{i,j} \xi_i Q^{-1}_{ij} \xi_j }
F(\xi_1,\ldots, \xi_M ) }
{\int \prod_{i=1}^M d^d \xi_i
\e^{-\frac 12 \sum_{i,j} \xi_i Q^{-1}_{ij} \xi_j }} \,.
\label{4.2}
\ee

The solution~\rf{4.1}, \rf{4.2} holds for an arbitrary $Q_{ij}$. For the 
choice~\rf{Qq} the inverse matrix can easily be calculated and the exponent
in \eq{4.2} reads
\be
\sum_{i,j} \xi_i Q^{-1}_{ij} \xi_j = \sum_{i=1}^M \left\{
\frac{(1-q)}{(1+q)} \xi_i^2 + \frac{q}{(1-q^2)} (\xi_i-\xi_{i-1})^2
\right\} \,.
\label{4.3}
\ee

The following comments concerning this formula are in order:
\vspace{-10pt}
\begin{itemize}
\addtolength{\itemsep}{-8pt}

\item[i)] The r.h.s.\ of \eq{4.3} is positively definite so that all integrals over 
$\xi_i$ in \eq{4.2} are convergent. 

\item[ii)] 
The inverse matrix $Q_{ij}^{-1}$ is local for the choice~\rf{Qq} --- 
only diagonal and next to diagonal matrix elements are nonvanishing.

\item[iii)] 
Typical loops which contribute to the average~\rf{4.2} are continuous but not 
smooth as $M\ra\infty$ --- one easily estimates that 
$(\xi_i-\xi_{i-1})^2 \sim (1-q)$ so that the typical loops are fractal.
\end{itemize}
\vspace{-9pt}

The property {iii)} can be reformulated in a slightly different way.
Let us consider the correlator of $\xi_i$ and $\xi_j$ which equals
\be
\LA \xi_{i}^{\mu} \xi_{j}^{\nu} \RA^{(M)}_\xi = \delta^{\mu\nu}\, Q_{ij} =
\delta^{\mu\nu}\, q^{|i-j|}
\label{4.4}
\ee
where $|i-j|$ is calculated again \ mod$\,M\,$. One sees that $\xi_i$ and 
$\xi_j$ are correlated for $|i-j|\la (1-q)^{-1}$ when $q\ra1$. Since $1\ll 
(1-q)^{-1}\ll M$, this implies that the continuum limit exists.  For $q=0$ 
which were correspond to the approximation of the functional Laplacian by means 
of the ordinary $d\cdot M$-dimensional Laplacian~\cite{Lev51}, $\xi_i$ and 
$\xi_j$ would be not correlated so that there would be no continuum limit.  

We can see now once more that our choice of $Q_{ij}$ satisfies a very important 
requirement: the terms of order ${\cal O}(M^{-1})$ which depend, in particular,
on details of the discretization of the averaging functional are not essential 
for smooth contours. This were not be the case, generally speaking, for other 
discretizations. For instance, the average of the quantity 
$(\xi_i-\xi_{i-1})^2$, which is ${\cal O}(M^{-2})$ for smooth contours, equals
(using \eq{4.4}) 
\be
\LA (\xi_i-\xi_{i-1})^2 \RA _{\xi}^{(M)} =2d\,(1-q) 
\label{4.11}
\ee
and would become ${\cal O}(1)$ if $q$ were not tend to $1$.
We see that the continuum limit exists only 
for $q\ra1$ when the r.h.s.\ of \eq{4.11} is vanishing.

\subsubsection{Remark on normal coordinates}

It is instructive to diagonalize the matrix $Q_{ij}^{-1}$ given by \eq{4.3}. 
This can always be 
done by a unitary transformation
\be
Q_{ij}^{-1} \ra \Omega_{ni} Q_{ij}^{-1} \Omega_{jl}^\dagger
= \l_n^{-1} \delta_{nl}
\ee
with
\be
\Omega_{nl} = \frac{1}{\sqrt{M}} \e^{2\pi i \frac{nl}{M}} 
\ee
and diagonal elements
\be
\l_n^{-1} =\frac{1+q^2-2q\cos{\frac{2\pi n}{M}}}{1-q^2}
~~~~~~(n=1,\ldots, M ~~~~~~\hbox{mod}\, M )\,.
\label{diagonal}
\ee
We have denoted the eigenvalues of the inverse matrix $Q_{ij}^{-1}$ by 
$\l^{-1}_n$ in order to keep the notation $\l_n$ for the eigenvalues of the 
matrix $Q_{ij}$ itself. As is seen from \eq{diagonal}, all $\l_n$ are positive
for $q\leq1$.

The diagonalization of $Q_{ij}$ can be viewed as a transition from the 
coordinates $\xi_i$ to the normal coordinates, $\alpha_n$, which are related by 
the Fourier transform
\be
\alpha_{n}^{\mu} =\frac 1M \sum_{j=1}^M \xi_{j}^{\mu}\,\e^{2\pi i \frac{jn}{M}}
\,,~~~~~~\alpha_{M-n}^{\mu}=(\alpha_{n}^{\mu})^* \,. 
\label{normal}
\ee
For translationally invariant functionals, nothing depends on 
$\alpha_M=\alpha_0$ which
is associated with a translation of the contour as a whole. For an even $M$ the 
coordinate $\alpha_{M/2}$ is real. For this reason the formulas below are 
valid, strictly speaking, for an odd $M$ and should be slightly modified for an 
even $M$.

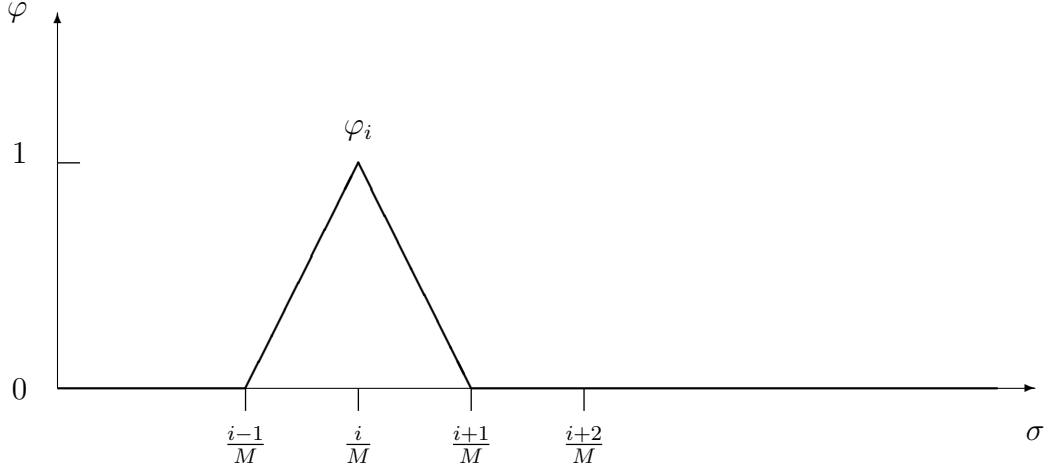
\begin{figure}[tbp]
\unitlength=1.00mm
\begin{picture}(44.0,63.00)(-10,70)
\thinlines
\put(10.00,80.00){\vector(1,0){130.00}}
\put(10.00,80.00){\vector(0,1){50.00}}
\thicklines
\put(10.00,80.00){\line(1,0){25.00}}
\put(35.00,80.00){\line(1,2){15.00}}
\put(50.00,110.00){\line(1,-2){15.00}}
\put(65.00,80.00){\line(1,0){70.00}}
\linethickness{0.5pt}
\put(10.00,110.00){\line(1,0){3.00}}
\put(35.00,80.00){\line(0,-1){3.00}}
\put(65.00,80.00){\line(0,-1){3.00}}
\put(50.00,80.00){\line(0,-1){3.00}}
\put(80.00,80.00){\line(0,-1){3.00}}
\put(50.00,113.00){\makebox(0,0)[cb]{$\varphi_i$}}
\put(6.00,130.00){\makebox(0,0)[rc]{$\varphi$}}
\put(6.00,110.00){\makebox(0,0)[rb]{$1$}}
\put(6.00,80.00){\makebox(0,0)[rc]{$0$}}
\put(35.00,75.00){\makebox(0,0)[ct]{$\frac{i-1}{M}$}}
\put(50.00,75.00){\makebox(0,0)[ct]{$\frac iM$}}
\put(65.00,75.00){\makebox(0,0)[ct]{$\frac {i+1}{M}$}}
\put(80.00,75.00){\makebox(0,0)[ct]{$\frac{i+2}{M}$}}
\put(140.00,75.00){\makebox(0,0)[ct]{$\sigma$}}
\end{picture}
\caption[x]   {\hspace{0.2cm}\parbox[t]{13cm}
{\small
   Simplest choice of the triangular basis functions given by 
   \eq{simplest}.}}
\label{Fig.5} 
\end{figure}

Notice that the normal coordinates~\rf{normal} can be related to
the standard normal coordinates in string theory:
\be 
\beta_k^\mu =  \int_0^1 d \s \xi^\mu(\s)
\e^{2\pi i \s k} \,,
\label{contnormal} 
\ee 
where $0\leq k \leq \infty$.
To establish the relation, let us substitute in the definition~\rf{contnormal}
$\xi(\s)$ given by the expansion~\rf{basis} with the simplest choice
\be
\varphi_i(\s) =\left\{ 
       \begin{array}{ll}
       1+(M\s-i)~~~ &  \hbox{for}~~\s \in [\frac{i-1}{M}, \frac{i}{M} ] \,, \\
       1-(M\s-i) &  \hbox{for}~~\s \in [\frac{i}{M}, \frac{i+1}{M} ] \,, \\
       0         &  \hbox{otherwise}
       \end{array}
       \right.
\label{simplest}
\ee
which is associated with $\s_i=i/M$ 
($0\leq \s\leq 1$) and the triangular shape.
The function~\rf{simplest} are depicted in Fig.~\ref{Fig.5}.
Doing the integral in \eq{contnormal}, one gets at finite $M$:
\be
\beta_k^\mu = 2\left( \frac{M}{2\pi k}\right)^2
\left(1-\cos{\fr{2\pi k}{M}}\right){\alpha}_k^\mu \,.
\label{contvsdisc}
\ee
We see from this formula 
that $\alpha_n$'s coincide for $n\ll M$, when the cosine can be expanded 
in the Taylor series, with the standard continuum normal coordinates 
$\beta_n$'s. They differ, however, for $n\sim M$.

For the normal coordinates~\rf{normal} the measure in \eq{4.2} factorizes and 
is pure Gaussian for each of $\alpha_n$'s:  
\be 
\prod_{i=1}^M d^d \xi_i 
\e^{-\frac 12 \sum_{i,j} \xi_i Q^{-1}_{ij} \xi_j } \propto
\prod_{n=1}^{[M/2]} \left(
\frac{d^d \re \alpha_n d^d \im \alpha_n}{(2\pi \l_n)^d }
\e^{-M\l^{-1}_n |\alpha_n|^2} \right) \,.
\label{distribution}
\ee
For an even $M$ there is no integration over $d^d \im \alpha_{M/2}$ since
$\alpha_{M/2} $ is real and the proper term in the exponent should be divided 
by $2$.

Since the Gaussian distribution~\rf{distribution} has dispersion 
$\sqrt{\l_n/2M}$ (or $\sqrt{\l_n/M}$ for $n=M/2$ at an even $M$),
the ratio of the dispersions for lower harmonics with $n\ll M$ and
 the higher harmonics with $n\sim M$ is proportional to $(1-q)^{-1}$.
Thus, the higher harmonics are suppressed which implies the continuity of the 
proper contours. 

\subsection{Some averages \label{ss3.5}}

The formula~\rf{4.4} and the Wick rules for calculating the Gaussian averages
allows one to derive some formulas for averaging the discretized functionals.
The useful identity 
\be
\BLA \xi_i^\mu F(\xi_{1},\ldots,\xi_{M}) \BRA_\xi^{(M)} =
\sum_j Q_{ij} \LA \frac{\d}{\d \xi_j^\mu }
F(\xi_{1},\ldots,\xi_{M}) \RA_\xi^{(M)} \,,
\label{discem}
\ee
which follows from the quadratic character of the action~\rf{4.3},
simplifies the calculation of averages of polynomial in $\xi_i$ expressions.

Further simplifications occur for the case when averaging 
expressions are just functions rather than discretized functionals.
If the averaging expression is a function of one of the variables $\xi_i$,
we get
\be
\BLA f(\xi_{i}) \BRA_\xi^{(M)} = \int \frac{d^d \xi}{(2\pi)^{\frac d2}}
\e^{-\frac{\xi^2}{2}} f(\xi)
\label{4.5}
\ee
which coincides~\cite{Lev51} with the formula for the average at $q=0$.

The average of a function of two variables reads
\be
\BLA f(\xi_{i},\xi_j) \BRA_\xi^{(M)} 
= \int \frac{d^d \xi_1}{(2\pi)^{\frac d2}}
\frac{d^d \xi_2}{(2\pi)^{\frac d2}}
\e^{-\frac{\xi_1^2+\xi_2^2}{2(1-G^2)}+G\frac{\xi_1\xi_2}{(1-G^2)}} 
f(\xi_1,\xi_2) (1-G^2)^{-\frac d2}
\label{4.6}
\ee
where $G\equiv q^{|i-j|}$ stands for the correlator~\rf{4.4}.

The extension of Eqs.~\rf{4.5}, \rf{4.6} to a function which depends on
$m$ variables $\xi_{i_1},\ldots,\xi_{i_m}$ reads
\be
\BLA f(\xi_{i_1},\ldots,\xi_{i_m}) \BRA_\xi^{(M)} = 
\prod_{j=1}^m \left(\int \frac{d^d \xi_j}{(2\pi)^{\frac d2}} \right)  
\,\e^{-\frac 12 \sum_{k,l=1}^m \xi_k G_{kl}^{-1} \xi_l}
f(\xi_1,\ldots,\xi_m) ( \det{G_{kl}})^{-\frac d2}
\label{A.I}
\ee
where $G_{kl}\equiv q^{|i_k-i_l|}$. The meaning of this formula is very simple: 
if the averaging function does not depend on the part of variables $\xi_i$, 
those can be integrated out in the general formula~\rf{4.2} with the result 
given by~\eq{A.I}. Similarly, the number of integration is decreased when the 
integrand, $f$, in \eq{A.I} does not depend on the whole set   
$\xi_{i_1},\ldots,\xi_{i_m}$ but depends only on some $n<m$ their combinations.

Let us illustrate the last statement by the $m=2$ example. Introducing in 
\eq{4.6} the new variables
\be
\eta_\pm=\frac{\xi_1+\xi_2}{\sqrt{2(1\pm G)}} \,,
\ee
we represent it as
\be
\BLA f(\xi_{i},\xi_j) \BRA_\xi^{(M)} = \int \frac{d^d \eta_+}{(2\pi)^{\frac d2}}
\frac{d^d \eta_-}{(2\pi)^{\frac d2}}
\e^{-\frac{\eta_+^2+\eta_-^2}{2}} 
f\left(\sqrt{\fr{1+G}{2}}\eta_+ + \sqrt{\fr{1-G}{2}}\eta_- ,
\sqrt{\fr{1+G}{2}}\eta_+ - \sqrt{\fr{1-G}{2}}\eta_- \right) \!.
\label{A.2}
\ee
Now if $f$ depends only on the combination $a \xi _i + b \xi_j$,
one gets the standard formula for 
averaging the independent Gaussian random quantities:
\be
\BLA f(a\xi_{i}+b\xi_j) \BRA_\xi^{(M)} = \int \frac{d^d \xi}{(2\pi)^{\frac d2}}
\e^{-\frac{\xi^2}{2}} f\left(\sqrt{a^2+b^2+2Gab}\,\xi\right) \,.
\label{A.3}
\ee
Substituting here $j=i-1$, $a=-b=1$ and $G=q$, one gets, in particular,
\be
\BLA f(\xi_{i}-\xi_{i-1}) \BRA_\xi^{(M)} = 
\int \frac{d^d \xi}{(2\pi)^{\frac d2}}
\e^{-\frac{\xi^2}{2}} f\left(\sqrt{2(1-q)}\,\xi\right) 
\label{A-3}
\ee
which recovers \eq{4.11}.

It is instructive to discuss how \eq{A.I} simplifies if one index, say $i_m$, 
is such that $|i_m-i_j| \gg (1-q)^{-1}$ ($j=1,\ldots,m$--$1$). Then 
correlations between $\xi_{i_m}$ and $\xi_{i_j}$ vanish,  $G_{jm}=0$ and the
measure in~\rf{A.I} factorizes. In particular, if all $|i_k-i_l| \gg 
(1-q)^{-1}$ ($k\neq l$), one gets 
\be 
\BLA f(\xi_{i_1},\ldots,\xi_{i_m}) \BRA_\xi^{(M)}
= \prod_{j=1}^m \left(\int \frac{d^d \xi_j}{(2\pi)^{\frac d2}} \,\e^{-\frac 
{\xi_j^2}{2}} \right) f(\xi_1,\ldots,\xi_m) \,.  
\label{4.12} 
\ee 
This formula 
is typical for averages over independent random quantities and coincides with 
that of Gateaux~\cite{Lev51} for an average over a sphere in $L_2$.

On the contrary, if all $|i_k-i_l| \ll (1-q)^{-1}$ ($k\neq l$), 
$G_{i_ki_l}=1$ and $\xi_{i_k}$ and $\xi_{i_l}$ are correlated. In this limit 
\eq{A.I} yields
\be
\BLA f(\xi_{i_1},\ldots,\xi_{i_m}) \BRA_\xi^{(M)}
= \int \frac{d^d \xi}{(2\pi)^{\frac 
d2}} \e^{-\frac{\xi^2}{2}} f(\xi,\ldots,\xi) 
\label{4.7} 
\ee
so that the Gaussian random quantities are totally dependent.

\subsection{The continuum limit \label{ss3.6}}

As is already mentioned above, a continuum limit of
the discretized formulas exists for $M\ra\infty$,\/
$q\ra 1$\/ and $M(1-q)\ra
\infty$\/ simultaneously. The simplest way to derive the proper continuum
 formulas is to use the normal coordinates which are introduced in
Subsect.~\ref{ss3.4} where it is shown that only the lower modes with 
$n\ll M$ are essential for $M\ra\infty$.

Let us expand the associated eigenvalues~\rf{diagonal} in $1/M$ for $n\ll M$:
\be
\l_n^{-1} \ra \frac{1-q}{2} -\frac{1}{2(1-q)}\left( \frac{2\pi n}{M}\right)^2 
\,.  
\ee
We now take the limit~\rf{3.3} in two steps: 1) $M\ra\infty$ at some
small but fixed
\be
\epsilon = \frac{1}{M(1-q)}
\label{defeps}
\ee 
and 2) $\epsilon \ra 0$. Nice continuum formulas will be written after the step 1) 
when $M \l^{-1}_n$, which enter the exponent in \eq{distribution}, are finite
\be
M \l_n^{-1} \ra \frac{1}{2\epsilon} -\frac{\epsilon}{2}\left( 2\pi n \right)^2 
\,.  
\label{contdiagonal} 
\ee

These eigenvalues coincide with the ones for an (Euclidean) harmonic 
oscillator with the frequency $\om =1/2$ at finite temperature $T = \epsilon /2$
which is described by the action
\be 
S = {1\over 2} \int_0^{\frac 2\epsilon} d t \left\{  
\left( \frac{d \xi(t)}{d t} \right)^2 + \frac 14 \xi^{2}(t ) \right\} \,.  
\label{8pr} 
\ee 
Therefore, the continuum limit of the action~\rf{4.3} is~\cite{Mak88}%
\footnote{As $\epsilon\to0$ this tends to  the averaging over the sphere in $L_2$ introduced by \label{foo3}
the french mathematician Ren\' e  G\^ ateaux~\cite{Gat37} (killed at World War I).
The presence of small but finite $\epsilon$ 
guarantees the continuity which is required for lines of force in Gauge Theory.}
\be 
S = {1\over 4} \int_0^1 d\sigma \left\{ \epsilon \dot\xi^{2}(\sigma ) + 
{1\over \epsilon } \xi ^{2}(\sigma ) \right\} \,.
\label{8}
\ee 
The variables $t$ ($0\leq t\leq 2/\epsilon$) and $\s$ ($0\leq \s\leq 1$) 
in~\rf{8pr} and \rf{8} are related by $\s=\epsilon t/2$. We prefer the 
variable $\s$ to the variable $t$ in order to have a finite interval 
($[0,1]$ in the given case) in analogy with string theory.

The continuum action~\rf{8} can alternatively be derived directly from~\rf{4.3}
using the formulas
\be
\int_0^1 d \s \xi^2(\s) = \frac 1M \sum_{i=1}^M \left( \frac{2\xi_i^2}{3}
+ \frac{\xi_i\xi_{i-1}
}{3}\right)
\ee
and
\be
\int_0^1 d \s \dot \xi^2(\s) = M \sum_{i=1}^M \left( 2\xi_i^2
- 2 \xi_i\xi_{i-1} \right)
\ee
which can be obtained by substituting the expansion~\rf{basis} with 
$\varphi_i(\s)$ given by \eq{simplest} on the l.h.s.\ and calculating 
the integrals. The continuum action~\rf{8} is recovered 
from the discretized one~\rf{4.3} as $M\ra\infty$ at fixed 
$\epsilon$ given by \eq{defeps}.

The continuum path-integral representations of
the $M\ra\infty$ limit of the above discrete formulas will be 
considered in the next section.

\newsection{Properties of loop-space Laplace equation\label{s3}}

The path-integral representation for the Green function of the 
loop-space Laplacian, which is constructed in the previous 
section taking the $M\ra\infty$ limit of the polygon discretization 
of loop space, can be alternatively derived directly in the continuum theory 
making a smearing of the functional Laplacian. The smearing is characterized by 
a parameter $\epsilon$ so that the loop-space Laplacian is recovered as $\epsilon\ra0$.
The Green function of the smeared operator is given by the Gaussian 
path-integral while the simplest exponential smearing leads to the action of 
the Euclidean harmonic oscillator at finite temperature.
The averages w.r.t.\ the Gaussian measure
can be calculated, including the case when  
velocities enter the averaging expression.
While the reparametrization invariance is explicitly broken at finite $\epsilon$, 
it can be shown that it restores for the 
reparametrization-invariant averaging expression as $\epsilon\ra0$.

\subsection{Smeared loop-space Laplacian}

Let us start from the relation between the partial derivatives
w.r.t.\ to the position of $i$-th vertex of the $M$-gon, $\d / \d x^\mu_i$,
and the continuum variational derivative, $\delta /\delta x^\mu(\s)$.
Using \eq{basis} and the chain rule, one gets
\be
{\d \over \d x^\mu_i}= \int_0^1 d\s \varphi_i(\s) 
{\delta \over \delta x^\mu(\s)}
\label{7.7}
\ee
where we choose $0\leq\s\leq 1$.
The meaning of this formula is that we look at $W$ by a left eye as at a 
function of the polygon vertices while by the right eye as at a functional of 
the whole polygon which is given by \eq{basis}. Similarly, one gets
for the operator~\rf{discDelta}
\be
\Delta^{(M)} = \int_0^1 d\s \int_0^1 d\s^\p 
\sum_{i,j=1}^M \varphi_i(\s) Q_{ij} \, \varphi_j(\s^\p)
{\delta \over \delta x^\mu(\s^\p)}{\delta \over \delta 
x^\mu(\s)}  \,.
\label{7.8}
\ee

Let us denote
\be
G(\s,\s^\p) \equiv \sum_{i,j=1}^M \varphi_i(\s) Q_{ij} \, \varphi_j(\s^\p) \,.
\label{defG}
\ee
If $\s$ and $\s^\p$ belong to the $i$-th and $i$+$k$-th intervals, 
respectively, one gets
\be
G(\s,\s^\p) = Q_k + (Q_{k+1}-Q_k)\left[
 \varphi_i(\s) + \varphi_{i+k}(\s^\p)  -2 \varphi_i(\s) \varphi_{i+k}(\s^\p) 
 \right]
\label{7.10}
\ee
for the general matrix~\rf{3.2}. If $(Q_{k+1}-Q_k)\ra 0$ as 
$M\ra\infty$ as it holds for~\rf{Qq}, the second term on the r.h.s.\ of
\eq{7.10} vanishes.  If in addition $\s_i=i/M$, then for $Q_k=q^k$ we find
\be
G(\s,\s^\p) = \e^{-\frac{|\s-\s^\p|}{\epsilon}}~~~~~~~(\epsilon \ll 1)
\label{Geps}
\ee
with 
\be
\eps=-\frac {1}{M \log{q}}
\ee
which coincides with~\rf{defeps} as $q\ra 1$. The function~\rf{Geps} coincides
in turn with the (Euclidean) Green function for the continuum action~\rf{8}%
\footnote{~The finite-temperature effects which are due to the periodicity, 
$\xi(0)=\xi(1)$, are negligible for $\eps\ll 1$.}.

For an arbitrary choice of $\s_i$ and $\varphi_i(\s)$ we obtain from \eq{7.10}
an arbitrary smearing function, $G(\s,\s^\p)$, which satisfies
\be
G(\s,\s) = 1;~~~ G(\s,\s+\h) \rightarrow  0  \hbox{ \ \ \ \ for \ }\h \neq  0
~~~\hbox{as}~~\epsilon\ra0~.
\label{2}
\ee
For $\epsilon\ll1$ this $G(\s,\s^\p)$ is related to~\rf{Geps} by a 
reparametrization.  

A general smearing function $G(\sigma ,\sigma ^\p)$ which obeys~\rf{2}
is associated with a generic smeared functional Laplacian
\be
\Delta^{(G)} =\int_0^1 d\sigma \int_0^1 d\sigma ^\prime  G(\sigma ,\sigma ^\p)
 {\delta \over \delta x_{\mu }(\sigma ^\prime )} {\delta \over \delta x_{\mu 
}(\sigma )} \,.
\label{2.13}
\ee
Separating the contribution from $\s=\s^\p$ which is associated according to 
\eq{defDelta} with the functional Laplacian $\Delta$, \eq{2.13} can be 
rewritten as
\be
\Delta^{(G)} = \Delta+\int_0^1 d\sigma \pintab d\sigma ^\prime  G(\sigma ,\sigma ^\p)
 {\delta \over \delta x_{\mu }(\sigma ^\prime )} {\delta \over \delta x_{\mu 
}(\sigma )}  \,.
\label{2.14}
\ee

Some comments concerning the smearing~\rf{2.13} are in order:
\vspace{-10pt}
\begin{itemize}
\addtolength{\itemsep}{-8pt}
\item[i)] 
The second term on the r.h.s.\ of \eq{2.14}, which involves the principal-value 
integral, is an addition to the functional Laplacian. It is an operator of 
second order (does not satisfy the Leibnitz rule~\rf{Leibnitz})  and is 
reparametrization noninvariant. 

\item[ii)]
The contribution of the  second  term is of order 
$\eps$ for smooth contours because the domain $|\s-\s^\p|\sim\eps$ is essential 
in the integral over $d\s^\p$. Therefore, when $\eps\ra0$, it is negligible, 
reparametrization invariance is restored and $\Delta^{(G)}$ tends to the 
functional Laplacian $\Delta$.

\item[iii)]
The smearing~\rf{2.13} is a special case of a more general smearing proposed by 
Migdal~\cite{Mig86}. \sloppy
Such extensions of the functional Laplacian are known in 
mathematics as infinite-dimensional elliptic operators of the L\'evy 
type~\cite{Fel86}.

\end{itemize}
\vspace{-9pt}

Equation~\rf{Geps} for the smearing function is written for a particular parametrization,
say the proper-length parametrization \rf{properlength}. Its reparametrization-invariant
extension is given by
\be
G(\s,\s')=\e^{-\big|\int_\s^{\s'} \!d\tau \sqrt{\dot x^2(\tau)}\big|/\eps}\qquad
(\eps\ll L),
\label{Gepsinv}
\ee
where $\eps$ has the dimension of length. 
It reproduces \rf{Geps} for the proper-length parametrization \rf{properlength}.

\subsection{Continuum Green function}

The operator $\Delta ^{(G)}$ which is defined by \eq{2.13}
can be inverted so that the equation
\be
\Delta ^{(G)} W [x] = J [x]
\label{3}
\ee
with the proper choice of boundary conditions can be solved to give
\be
W [x] = 1 - {1\over 2}\int_0^\infty dA \left\{\LA J [x+\sqrt{A}\xi ] 
\RA_{\xi}^{(G)}
- \LA J [\sqrt{A}\xi ] \RA_{\xi }^{(G)} \right\}\, .
\label{4}
\ee
Here the average over the loops $\xi(\s)$ is given by the path integral
\be
\BLA F [\xi] \BRA_\xi^{(G)} = \frac{\int_{\xi(0)=\xi(1)} D\xi \e^{-S} F [\xi]}
{\int_{\xi(0)=\xi(1)} D\xi \e^{-S}}
\label{5}
\ee
and  
\be 
S = {1\over 2}\int_0^1 d\sigma 
\int_0^1 d\sigma ^\prime \left\{ \xi (\sigma ) G^{-1}(\sigma, 
\sigma ^\prime ) \xi (\sigma ^\prime )\right\} 
\label{6} 
\ee 
where $G^{-1}(\s,\s')$ stands for the inverse operator.
For the simplest 
case~\rf{Geps} one gets the local action~\rf{8}, while for the invariant smearing~\rf{Gepsinv}
one shifts
\be
\epsilon \to \frac \eps {\sqrt{\dot x^2(\s) }}
\label{epsshift}
\ee
which
results in the reparametrization-invariant local action
\be 
S = {1\over 4} \int_0^1 d\sigma \left\{ \frac \eps {\sqrt{\dot x^2(\s) }} \dot\xi^{2}(\sigma ) + 
\frac  {\sqrt{\dot x^2(\s) }}\eps\xi ^{2}(\sigma ) \right\} \,.
\label{8inv}
\ee

The formulas~\rf{4} to \rf{6} as well as other continuum formulas can be 
derived using the ones for the polygon discretization as is explained in 
Subsect.~\ref{ss3.6}.
We shall present an alternative derivation directly using the continuum 
smearing~\rf{2.13} without any reference to the polygon discretization.

In order to prove Eqs.~\rf{4} to \rf{6}, let us consider~\rf{5} as a partition 
function of the oscillator system with the action~\rf{6} and derive  
the (integrated over $d \s^\p$) quantum equation of motion:  
\be 
\BLA { 
\xi}^{\mu }(\sigma ) F[\xi ] \BRA_{\xi } ^{(G)} = \int^{1}_{0} d\sigma^\prime { 
G}(\sigma -\sigma ^\prime ) \LA {\delta  F[\xi ]\over \delta \xi ^\mu (\sigma 
^\prime )} \RA_{\xi}^{(G)}\,, 
\label{em} 
\ee 
which results from the invariance 
of the measure $D\xi$ under an arbitrary shift \be \xi^\mu(\s) \ra \xi^\mu(\s) 
+ \delta \xi^\mu(\s)~,~~~~~~ \delta \xi^\mu(0)=\delta \xi^\mu(1) \label{shift} 
\ee
and $F[\xi]$ is an arbitrary functional. \eq{em} is the continuum analogue of 
the formula~\rf{discem} for the discrete case.

To derive \eq{em} as a (quantum) equation of motion for the local action~\rf{8inv}, let us  
make the shift~\rf{shift}, that gives
\be
\frac 12 \LA  \left\{ -\frac d{d\s} \frac \eps {\sqrt{\dot x^2(\s) }} 
 \dot \xi^\mu(\s)+\frac  {\sqrt{\dot x^2(\s) }}\eps 
\xi^{\mu }(\sigma ) \right\} F[\xi ] \RA_{\xi } ^{(G)}
= 
\LA {\delta  F[\xi ]\over \delta \xi ^\mu (\sigma )} \RA_{\xi}^{(G)}
\label{8eminv}
\ee
which is nothing but the usual set of Schwinger--Dyson equations for the 
harmonic oscillator. 
The standard equation%
\footnote{~Due to the periodicity, $\delta(\s-\s^\p)$ on the r.h.s.\ is the
periodic delta-function with the period $1$.}
\be
-\frac d{d\s} \frac \eps {\sqrt{\dot x^2(\s) }}  \dot G(\s-\s^\p)+
\frac  {\sqrt{\dot x^2(\s) }}\eps G(\s-\s^\p) = 2 \delta(\s-\s^\p)
\label{SDDinva}
\ee
satisfied by the two-point Green function~\rf{Gepsinv} can be obtained 
from  \eq{8eminv} substituting $F[\xi]=\xi^\nu(\s)$.
\eq{em} can be derived from \eq{8eminv} multiplying by $G(\s,\s^\p)$ 
and integrating over $\s^\p$.

The proof of Eqs.~\rf{4} to {\rf{6} is based on the following sequence of 
formulas. Using the chain rule, one gets
\be
\frac{2d}{dA}J[x+\sqrt{A}\xi] = \int_0^1 d\s \frac{\xi^\mu(\s)}{A}
\frac{\delta}{\delta \xi^\mu(\s)} J[x+\sqrt{A}\xi]\,.
\ee
Averaging over $\xi$ and using the equation of motion~\rf{em} and
the definition~\rf{2.13} of $\Delta^{(G)}$, we get
\bea
\frac{2d}{dA}\LA J[x+\sqrt{A}\xi] \RA_\xi^{(G)} &=& 
\int_0^1 d\s \int_0^1 d\s^\p
G(\s-\s^\p) \LA \frac{\delta^2 J[x+\sqrt{A}\xi]}{\delta x^\mu(\s^\p)
\delta x^\mu(\s)} \RA_\xi^{(G)} \non & =& 
\Delta^{(G)} \LA J[x+\sqrt{A}\xi] \RA_\xi^{(G)} \,.
\label{2.23}
\eea
One can apply now $\Delta^{(G)}$ to \eq{4} and use~\rf{2.23} integrated 
over $A$. The result coincides with \eq{3} which completes the proof.

It is worth mentioning that the simple Gaussian formulas~\rf{5}, \rf{6}
for the Green function of the smeared Laplacian are due to 
presence of the second term on the r.h.s.\ of \eq{2.14}. 
Since this term is of second order in the variational derivative, the 
Green function is given by the path integral with the quadratic action.
While this would be no longer true in the limit $\eps=0$, all the formulas 
for averages we need can be calculated at finite (but small) $\eps$
taking then the limit $\eps\ra0$. 

\subsection{Some continuum averages \label{ss4.3}}

Since the path integral over $D\xi $ in \eq{5} is Gaussian, the main  formula 
for calculating the averages reads 
\be 
\BLA \xi^\mu (\sigma) \xi^\nu (\sigma^\p) \BRA_{\xi }^{(G)}
= \delta ^{\mu \nu } G(\sigma,\sigma^\p)\,.  
\label{9} 
\ee 

\eq{9} can be alternatively obtained by taking the $M\ra\infty$ limit 
of the polygon discretization. This is done by substituting the 
expansion~\rf{basis} for $\xi(\s)$ and $\xi(\s^\p)$ and calculating the 
discrete average according to \eq{4.4}. One gets finally
\be
\BLA \xi^\mu(\s) \xi^\nu(\s^\p) \BRA_\xi^{(M)} 
= \delta^{\mu\nu} \sum_{i,j=1}^M \varphi_i(\s) Q_{ij} \, \varphi_j(\s^\p)
\ee
which coincides with the r.h.s.\ of \eq{9} due to \eq{defG}.

Since \eq{9} looks like \eq{4.4} providing one denotes 
\be
G_{ij} \equiv G(\s_i,\s_j)\,,
\ee 
formulas for the continuum averages over $\xi$ of the functions 
$f\left(\xi(\s_1),\ldots,\xi(\s_m)\right)$ 
are identical to the formulas of Subsect.~\ref{ss3.5}. 
For instance, we have
\be
\BLA f\left(\xi(\s_1),\ldots,\xi(\s_m)\right) \BRA_\xi^{(G)} = 
\prod_{j=1}^m \left(\int \frac{d^d \xi_j}{(2\pi)^{\frac d2}} \right)
\,\e^{-\frac 12 \sum_{k,l=1}^m \xi_k G_{kl}^{-1} \xi_l}
f(\xi_1,\ldots,\xi_m) ( \det{G_{kl}})^{-\frac d2}
\label{3.5}
\ee
in full analogy with \eq{A.I}. The analogues of the other formulas~\rf{4.5} to 
\rf{A-3} can easily be written as well.

It is easy to obtain from~\eq{3.5} the following formula
\be 
\LA \e ^{i\sqrt{A} \sum_{j} p_j \xi (\sigma_j)} 
\RA_{\xi }^{(G)} = \e^{-{A\over 2} \sum_{i,j} p_i  G_{ij} p_j}
\label{10} 
\ee 
where $A$ is a constant and $p_j^{\mu }$ is some set of $d$-momenta.
\eq{10} will be used in the next section for an iterative 
solution of \eq{4} with the current $J[x]$  associated with large-$N$ QCD.

Let us consider now the limit of $\eps\ra0$ when the approximating 
operator~\rf{2.13} recovers the functional Laplacian.
Since $G_{kl}\ra0$ in this limit if $\s_k\neq\s_l$, one gets
\be
\lim_{\eps\ra0} \BLA f\left(\xi(\s_1),\ldots,\xi(\s_m)\right) \BRA_\xi^{(G)} =
\prod_{j=1}^m \left(\int \frac{d^d \xi_j}{(2\pi)^{\frac d2}}  
\,\e^{-\frac {\xi_j^2}{2} }\right)
f(\xi_1,\ldots,\xi_m) 
\label{Gateux}
\ee
like \eq{4.12} for independent Gaussian random quantities. 
Therefore, the 
average of a function which depends on a set of {\it different}\/ $\xi(\s_i)$ 
coincides at $\eps=0$ with the Gateaux averages over the sphere in 
$L_2$ (see footnote~\ref{foo3}). 
For this reason there were no need in the polygon discretization and/or the 
above smearing procedure for these averages. 

In gauge theories one deals, however, with functionals which are 
integrals of functions depending explicitly on the velocity 
$\dot{\xi}^\mu(\s)$ like the Wilson loop~\rf{Wl}. 
Such averages which can not be managed within the Gateaux approach are 
considered in the next subsection. 

\subsection{Averages with velocity \label{ss4.4}}

Let us start from the simplest average
\be
\LA \dot{\xi}^\mu (\sigma) \xi^\nu (\sigma^\p) \RA_{\xi }^{(G)}
= \delta ^{\mu \nu } \dot  G(\sigma,\sigma^\p), \qquad \dot  G(\sigma,\sigma^\p)\equiv
 \frac d{d\s} G(\sigma,\sigma^\p)
\label{dot9} 
\ee 
which includes both $\xi^\nu(\s^\p)$ and the velocity $\dot{\xi}^\mu(\s)$.
This formula is obtained differentiating \eq{9} w.r.t.\ $\s$.
It is important that the correlator of $\dot{\xi}^\mu(\s)$ and $\xi^\nu(\s)$ 
vanishes since 
\be
\dot G(\sigma,\sigma^\p)= \hbox{sign}(\s'-\s) \frac{\sqrt{\dot x^2(\s)}}\eps G(\s,\s')
\stackrel{\s'=\s}=0.
\ee

The averages which include both a function of $\xi^\nu(\s_i)$ and
the velocity $\dot{\xi}^\mu(\s)$ can be
derived differentiating \eq{em} w.r.t.\ $\s$ which yields
\be
\LA\dot{ \xi}^{\mu }(\sigma ) F[\xi ] \RA_{\xi } ^{(G)}
= \int^{1}_{0} d\sigma^\prime
\dot{ G}(\sigma ,\sigma ^\prime )
\LA {\delta  F[\xi ]\over \delta \xi^\mu (\sigma ^\prime )} \RA_{\xi}^{(G)}.
\label{14}
\ee
If $F[\xi]=f\left(\xi(\s_1),\ldots,\xi(\s_m)\right)$, one gets
\be
\LA\dot{ \xi}^{\mu }(\s) f\left(\xi(\s_1),\ldots,\xi(\s_m)\right)  
\RA_{\xi } ^{(G)} =
\sum_{i=1}^m \dot G(\s,\s_i)
\LA\frac{\d}{\d \xi^\mu(\s_i)} 
f\left(\xi(\s_1),\ldots,\xi(\s_m)\right)  \RA_{\xi } ^{(G)}
\label{velocity}
\ee
The average on the r.h.s.\ is calculable  as $\eps\ra0$ with the aid of 
\eq{Gateux}. If some of $\s_i$ coincides with $\s$, the proper term in the sum
over $i$ should be omitted because $\dot G(\s,\s)=0$.

If the function $f$ is the exponential like in \eq{10}, one gets 
\be 
\LA  \dot \xi ^{\mu }(\sigma _k) \e^{i\sqrt{A} 
\sum_{j}p_j \xi (\sigma_j)} \RA_{\xi }^{(G)} = i \sqrt{A}
\sum_{m}\dot{ G}_{km} p_m^{\mu } 
\e^{-{A\over 2} \sum_{i,j} p_i G_{ij} p_j}
\label{15} 
\ee 
where $\dot G_{km}$ stands for $\dot G(\s_k,\s_m)$.
This formula is 
used below at iterative solution of \eq{4}.

It is instructive to consider in some detail the case when the averaging 
expression is the contour integral
\be
\int_{\xi_0}^{\xi_f} d \xi^\mu f(\xi_f,\xi) \equiv 
\int_{\s_0}^{\s_f} d\s \dot{\xi}^\mu(\s) f(\xi(\s_f),\xi(\s))
\label{3.14}
\ee
where the integration runs along the open contour from the point 
$\xi_0=\xi(\s_0)$ to the point  $\xi_f=\xi(\s_f)$ and the integrand 
depends explicitly on $\xi_f$. Using Eqs.~\rf{14} and \rf{3.5}
with $m=2$, we get
\bea
\lefteqn{\LA \int_{\xi_0}^{\xi_f} d \xi^\mu f(\xi_f,\xi) \RA_{\xi }^{(G)} 
= \int_{\s_0}^{\s_f} d\s \dot G(\s,\s_f)
\LA\frac{\d}{\d \xi^\mu(\s_f)} f(\s_{f},\s )  \RA_{\xi } ^{(G)}}
\non
&&= \int_0^1 d G (1-G^2)^{-\frac d2} 
 \int \frac{d^d \xi_1 
\,d^d \xi_2}{(2\pi)^{d}}
\e^{-\frac{\xi_1^2+\xi_2^2}{2(1-G^2)}+G\frac{\xi_1\xi_2}{(1-G^2)}} 
\frac{\d}{\d \xi_1^\mu} f(\xi_1,\xi_2) \,.
\label{3.15}
\eea
The explicit form of $G$ is not essential in the derivation. 
It was used only the property~\rf{2}. 
This is because of the reparametrization invariance of the averaging 
expression.  The reason why the dependence on $G$ is left in \eq{3.15} even for 
$\eps\ra0$ is as follows. The typical region of $\s$ which contribute to the 
integral over $d\s$ in \eq{3.15} is $|\s-\s_f| \sim \eps$ because $\dot G$ 
vanishes otherwise. However, the integral is not vanishing as $\eps\ra0$ since 
$\dot G\sim 1/\eps$ in this region. Therefore, the expression on the r.h.s.\ of 
\eq{3.15} arises as a result of doing an uncertainty of the type $\eps\times 
\eps^{-1}$.  It is the place where our method of smearing (and/or 
discretizing) the loop-space Laplacian works. There is impossible to perform 
this calculation within the framework of Ref.~\cite{Lev51}.

There are no extra complications to calculate the averages of more than one 
velocity $\dot\xi(\s_i)$. They can be calculated by virtue of \eq{14} where 
the functional $F[\xi]$ depends both on $\xi(\s_j)$ and $\dot\xi(\s_i)$.  
For the case of two velocities one gets
\bea
\lefteqn{\LA\dot{ \xi}^{\mu }(\sigma ) \dot{ \xi}^{\nu }(\sigma^\p )
F[\xi ] \RA_{\xi } ^{(G)} }
\non  &&=\delta^{\mu\nu}\frac d{d \s}\frac d{d \s'}G(\s,\s^\p) +
\int^{1}_{0} d\sigma_1
\int^{1}_{0} d\sigma_2
\dot{ G}(\sigma ,\sigma_1 )
\dot{ G}(\sigma^\p ,\sigma_2 )
\LA {\delta^2  F[\xi ]\over \delta \xi ^\mu (\sigma_1) \delta \xi ^\nu 
(\sigma_2)} \RA_{\xi}^{(G)}\!\!\!. ~~~~~~~
\label{dot14} 
\eea

The simplest example is 
\be 
\LA \dot{\xi}^\mu (\sigma) \dot{\xi}^\nu (\sigma^\p) 
\RA_{\xi }^{(G)} = \delta ^{\mu \nu } \frac d{d \s}\frac d{d \s'}
 G(\sigma,\sigma^\p) \label{ddot9} 
\ee 
which corresponds to $F[\xi]=1$ in \eq{dot14}. \eq{ddot9} can be easily 
verified differentiating \eq{dot9}.

For the exponential $G$ given by \eq{Gepsinv} when \eq{SDDinva} holds, one gets
on the r.h.s.\ of \eq{ddot9}                 
\be
\frac {d}{d\s} \frac {d}{d\s'} G(\s,\s^\p) = 
\frac {\sqrt{x^2(\s)}}{\eps} \frac {\sqrt{x^2(\s')}}{\eps}G(\s,\s^\p) - 2\frac {\sqrt{x^2(\s)}}{\eps} \delta(\s-\s^\p) \,.
\ee
While this function is divergent at $\s=\s^\p$, 
averaging expressions of such a type do 
not appear when averaging reparametrization-invariant functionals of the Stokes 
type. The result of averaging a Stokes functional is Stokes again.

Since Eqs.~\rf{velocity} and \rf{ddot9} are known, it is easy to calculate 
more complicated averages with two and three velocities:
\be
\LA   \dot \xi ^{\mu }(\sigma _k) 
\dot \xi ^{\nu }(\sigma _l)\e^{i\sqrt{A} 
\sum_{j}p_j \xi (\sigma_j)} \RA_{\xi }^{(G)} = 
- \left( A\sum_{m,n}\dot{ G}_{km} \dot{ G}_{ln} p_m^{\mu } p_n^{\nu }
+ \delta^{\mu\nu} \ddot{ G}_{kl} \right)
\e^{-{A\over 2} \sum_{i,j} p_i G_{ij} p_j}
\label{dot15}
\ee
and
\bea
\LA   \dot \xi ^{\mu }(\sigma _k)
\dot\xi ^{\nu }(\sigma _l)\dot\xi ^{\l }(\sigma _r)\e^{i\sqrt{A}
\sum_{j}p_j \xi (\sigma_j)} \RA_{\xi }^{(G)}  = -i \sqrt{A}
\left( A\sum_{m,n,s}\dot{ G}_{km} \dot{ G}_{ln} \dot{ G}_{rs}
p_m^{\mu } p_n^{\nu }p_s^\l \right. \non \left. - 
\sum_m \left( \delta^{\mu\nu} \ddot{ G}_{kl} \dot{ G}_{rm} p_m^\l +
\delta^{\mu\l} \ddot{ G}_{kr} \dot{ G}_{lm} p_m^\nu +
\delta^{\nu\l} \ddot{ G}_{lr} \dot{ G}_{km} p_m^\mu
\right) \right)
\e^{-{A\over 2} \sum_{i,j} p_i G_{ij} p_j} \,.
\label{ddot15}
\eea
These formulas will be used in the next section.

Notice that Eqs.~\rf{10}, \rf{15}, \rf{dot15}, \rf{ddot15}  as well as 
analogous formulas with more velocities can be deduced from the general formula 
\be
\LA \e ^{i\sqrt{A} \int d\sigma \dot p(\sigma )\xi (\sigma )} \RA_{\xi}^{(G)} 
= \e^{-{A\over 2} \int d\sigma \int d\sigma ^\prime \dot p(\sigma )G(\sigma,
\sigma ^\prime )\dot p(\sigma ^\prime )} 
\label{10'}
\ee 
where $p^{\mu }(\sigma )$ ($p^{\mu }(0)=p^{\mu }(1)$)  represents  a 
momentum-space loop~\cite{Mig86}. \eq{10'}  can  easily  be  proven  using 
\eq{9}. \eq{10} can be obtained from~\rf{10'}  by substituting a stepwise
$p^\mu(\s)$:
\be
\dot{p}^\mu(\s) = \sum_j p_j^\mu \delta(\s-\s_j),
\label{momloop}
\ee
while Eqs.~\rf{15}, \rf{dot15}, \rf{ddot15}, {\it etc.}\ can be derived from 
\eq{10'} first applying the proper number of times
the variational derivative $\delta /\delta 
p^{\mu}(\s_k)$  
according to the rule
\be
\dot {\xi}^\mu (\s_k) \ra \frac{i}{\sqrt{A}}
\frac{\delta }{\delta p^{\mu}(\s_k)} 
\ee 
and then substituting~\rf{momloop}.

\subsection{Remark on reparametrization invariance \label{ss4.5}}

The fact that the r.h.s.\ of \eq{3.15} does not depend on the form of
the smearing function $G$ is related, as is already mentioned, to the 
reparametrization invariance of the averaging expression~\rf{3.14}.  Let us 
prove the general property that the average~\rf{5} of any 
reparametrization invariant 
functional $F[\xi]$ is reparametrization invariant.

To this aim let us calculate
\bea
\lefteqn{\dot{x}^\mu(\s) \frac{\delta}{\delta x^\mu(\s)} \LA F[x+\sqrt{A}\xi]
\RA_\xi^{(G)} }\non&&= -
\dot{x}^\mu(\s)  \LA\frac{\delta S}{\delta x^\mu(\s)} F[x+\sqrt{A}\xi]
\RA_\xi^{(G)} -\LA \dot{\xi}^\mu(\s) \frac{\delta}{\delta \xi^\mu(\s)}  F[x+\sqrt{A}\xi]
\RA_\xi^{(G)}\!\! ,
\label{44}
\eea
where the first term on the r.h.s. appears because of the explicit dependence of the action~\rf{8inv}
on $\dot x^\mu(\s)$. Calculating the variation of the action, we obtain
\be
 -\dot{x}^\mu(\s) \frac{\delta S}{\delta x^\mu(\s)}=\frac 12 \dot \xi_\mu (\s)
 \left\{-\frac d {d\s} \frac\eps {\sqrt{ \dot x^2(\s)}} \dot\xi^\mu(\s)+
 \frac {\sqrt{ \dot x^2(\s)}}\eps \xi^\mu(\s)
 \right\}.
\ee
We can now use the Schwinger-Dyson equation~\rf{8eminv} to show that the first term on the
r.h.s. of \eq{44} is mutually canceled with the second term. This completes the proof.

It is worth noting once again that the cancellation of the two terms on the r.h.s. of \eq{44}
occurs because of an explicit dependence of the action~\rf{8inv} on $\dot x^\mu(\s)$.If we were rather use the action~\rf{8} which does not depend on $x^\mu(\sigma)$, 
the first term on the
r.h.s. of \eq{44} would be missing. Reparametrization invariance would then be restored only as
$\eps\to0$ for the following reason.

Let us rewrite the second term on the r.h.s. of \eq{44} using \eq{14} as
\be
\LA \xi_\mu(\s)\frac{\delta 
J[x+\sqrt{A}\xi]}{\delta \xi^\mu(\s)} \RA_\xi^{(G)} 
=  \int_0^1 d\s^\p \dot{G}(\s-\s^\p) \LA \frac{\delta^2 
J[x+\sqrt{A}\xi]}{\delta \xi^\mu(\s^\p) \delta \xi^\mu(\s)} \RA_\xi^{(G)} .
\label{2.24} 
\ee
Since the last expression vanishes as $\eps\ra0$ the reparametrization 
invariance is restored at $\eps=0$.

This vanishing occurs because the integral over $\s^\p$ on the r.h.s.\ of 
\eq{2.24} is not ordered, \ie while $|\s-\s^\p|\sim\eps$ both domains
$\s^\p<\s $
 and $\s^\p>\s$ contribute. These contribution comes with opposite signs since 
 $\dot G(\s-\s^\p)$ is an odd function. This was not the case for the ordered 
 integral over $\s^\p$ in \eq{3.15} where only one of two domains contributes.
 For this reason only averages of ordered contour integrals yield a nontrivial 
 contribution. There ordered contour integrals are typical for non-Abelian 
 gauge theories and, as we shall see in the next section, will lead to the 
 three-gluon vertex.
 
\subsection{Relation to loop operator in IIB model\label{s:regul} (added Nov.\ 2005)}


It is instructive to compare the smearing \rf{Geps}, \rf{Gepsinv} with the one of Ref.~\cite{IIB}
where the loop operator can be represented as
\be {\bbox \Delta}^{(G)} = \int d \s
\int_{\s-\epsilon}^{\s+\epsilon} d \s' \left( \frac{\delta^2}{\delta
x^\mu(\s')\delta x_\mu(\s)}+ \frac{\delta^2}{\delta Y^i(\s')\delta
Y_i(\s)}+ \frac{\delta^2}{\delta \zeta(\s')\delta \bar\zeta(\s)}\right),
\label{2.41eps}
\ee
by identifying $k^\mu(\s)$ with $\dot x^\mu(\s)$ and using the formula 
\be
\frac \delta {\delta x^\mu(\s)}= \frac d {d\s}\frac \delta {\delta \dot x^\mu(\s)}
\ee
of the variational calculus. The operator \rf{2.41eps} can be rewritten in 
 in the form~\rf{2.13} with
\be
G(\s,\s^\p)=\theta\Big( |\s-\s^\p|+\frac \epsilon 2\Big)-
\theta\Big( |\s-\s^\p|-\frac \epsilon 2\Big)
\ee
where $\theta$ is the Heaviside step function.

\subsection{Appendix to Sect.~\ref{s3}: Some useful formulas}

\subsubsection{$\Delta^{(G)}$ as an average}

The action of the (smeared) functional Laplacian on a functional $F[x]$ can be represented 
as the average
\be
\Delta^{(G)} F[x]=\frac {d^2}{d r^2}\BLA F[x+r \xi] \BRA_\xi^{(G)} \Big| _{r=0} .
\label{4.46}
\ee
This formula can be proven by expanding in $r$ and using \eq{9}
\bea
\BLA F[x+r \xi] \BRA_\xi^{(G)} &=&F[x]+ \frac{r^2}2 \int d\s_1 \int d\s_2 
\BLA \xi^\mu(\s_1)\xi^\nu(\s_2) \BRA_\xi^{(G)} 
\frac{\delta^2 F[x] }{\delta x^\mu(\s_1) \delta x^\nu(\s_2)} +{\cal O}(r^4) \non
&=&F[x]+ \frac{r^2}2 \int d\s_1 \int d\s_2 \,
G(\s_1,\s_2)
\frac{\delta^2 F[x] }{\delta x^\mu(\s_1) \delta x^\nu(\s_2)} +{\cal O}(r^4)
\label{4.47}
\eea
Equation~\rf{4.46} is more convenient for practical calculations than the definition~\rf{2.13}.

\subsubsection{Stokes' ansatz}

We shall need in the next section how $\Delta^{(G)}$ acts on the Stokes functionals
of the type
\be
F[x]=\oint d x_1^\mu \oint d x_2^\mu f(x_1-x_2), 
\label{4.52}
\ee
where $f$ is a function. Using \eq{4.46} we obtain
\bea
\Delta^{(G)} F[x]= 2 \int d\s_1\int d \s_2 \,\Big\{ \dot x^\mu(\s_1) \dot x^\mu(\s_2) 
\big[1-G(\s_1,\s_2)\big] \partial^2_\nu f\big(x(\s_1)-x(s_2)\big) \non
+(d-2) f\big(x(\s_1)-x(s_2)\big) \frac {d^2}{d s_1 d\s_2} G(s_1,\s_2) \Big\}.~~~
\label{4.53}
\eea
Note that the second term in the integrand vanishes in $d=2$. It comes from the action 
of the operator $\Delta^{(G)} $ on the velocities $\dot x^\mu (\s)$ (at least once).
The action of $\Delta^{(G)} $ on the function \mbox{$ f\big(x(\s_1)-x(s_2)\big)$} is given by
\be
\Delta^{(G)} f\big(x(\s_1)-x(s_2)\big)=2\big[1-G(\s_1,\s_2)\big] 
\partial^2_\nu f\big(x(\s_1)-x(s_2)\big)
\ee
which is of the type of the first term in \eq{4.53}.

The simplest diagram of perturbation theory for $W(C)$ has the form of \eq{4.52}
with $f$ being the free scalar propagator $D_0$. Then the first term in the integrand in \eq{4.53}
reproduces the r.h.s. of the loop equation to this order while the integral of the second term
is proportional to $(d-2) \eps L \Lambda^d$, where $\Lambda$ is a UV cutoff (of the dimension
of mass). It is negligible at $d>2$ only for $\eps \ll \Lambda^{-1}$ supplementing
the comment ii) after \eq{2.14}.

\newsection{Iterative solution and Feynman diagrams\label{s4}}

The loop equation of large-$N$ QCD can be solved by  iterations  in
the coupling constant, $\lambda $. To do this, one should calculate
averages w.r.t.\  the Gaussian measure using the formulas of the previous 
section. In contrast to Refs.~\cite{MM79,MM81} this does not assume any 
particular ansatz of the solution.
The iterative solution reproduces to order $\lambda $ the Feynman diagram with
gluon propagator and to order $\lambda ^{2}$ the whole set of 
diagrams for the Wilson loop average, including the one with the three-gluon 
vertex.  The iterative solution simplifies using some kind of a momentum-space 
representation which allows to pursue  
iterations, in principle, to any order in $\l$.   

\subsection{The iterative scheme}

For the case of large-$N$ QCD, the current 
$J[x]$ on the r.h.s.\  of  Eqs.~\rf{3} and \rf{4} 
is given due to \eq{Le} by
\be 
J[x] = \lambda \int_C 
dx_1^\mu \pintc dx_2^\mu \delta ^{(d)}(x_1- x_2) W (C_{x_1x_2}) W(C_{x_2x_1}).  
\label{11}
\ee 
This current depends on the functional $W$
bilinearly,  so  that  a non-linear equation for $W$ arises. We shall describe 
how this equation can be solved iteratively.

It is convenient to pass to  a  momentum-space  representation
for $W[x]$. The representation we use is not  the functional  Fourier
transformation to momentum-space loops advocated in Refs.~\cite{Mig86,BVM86}.
Instead, we expand loop functionals in a series of ordered  contour
integrals of the form
\be
F[x] = 1 +  \sum_{n=2}^\infty 
\int dx_1^{\mu_1}\ldots dx_n^{\mu_n} \theta_c (1\ldots n) 
F^{(n)}_{\mu _1\ldots \mu _n}(x_1,\ldots ,x_n) 
\label{12}
\ee
and then
simply perform a Fourier transformation of  the  integrand
\be
F^{(n)}_{\mu _1\ldots \mu _n}(x_1,\ldots ,x_n) =
\int  \prod_{i=1}^n \left({d^{d}p_i\over (2\pi )^{d}}\right)
\e^{i\sum_{j=1}^n p_j x_j} F^{(n)}_{\mu _1\ldots \mu_n}(p_1,\ldots, p_n ) \,.
\label{13}
\ee
The symbol $\theta_c (1\ldots n)$ stands for the contour theta-function which
orders the points $x_1,\ldots,x_n$ along the contour 
preserving the symmetry under cyclic 
permutations:
\bea
\lefteqn{
\int_0^1 d\s_1 \ldots \int_0^1 d \s_n \theta_c (1\ldots n) = 
\frac 1n \left\{ \int_0^1 d \s_1 \int_{\s_1}^1 d\s_2
 \ldots \int_{\s_{n-1}}^1 d \s_n \right. }\non &&+ \left.
 \int_0^1 d \s_2 \ldots  \int_{\s_{n-1}}^1 d \s_n  \int_{\s_n}^1 d\s_1
+\ldots + 
\int_0^1 d \s_n \int_{\s_{n}}^1 d \s_1 \ldots  \int_{\s_{n-2}}^1 d\s_{n-1}
\right\}\,.
\eea

After the substitution
$x(\sigma )\rightarrow x(\sigma )+\sqrt{A} \xi (\sigma )$
into  the  Ansatz~\rf{12}, \rf{13}, as is prescribed by Eq.~\rf{4},
a  variety
of  terms with different numbers of velocities,  $\dot\xi(\sigma_k)$,
arises.
Those can  be averaged over $\xi $ using the identities described in
Subsects.~\ref{ss4.3}, \ref{ss4.4}. The result is to be identified with the 
Feynman diagrams for $W[x]$.

As is shown in the previous section, the path-integral in the integral 
version~\rf{4} of the loop equation is Gaussian at finite $\eps$.
For the reason which becomes clear from the next subsection, we 
replace~\rf{11} at finite $\eps$ by
\bea
J^{(G)}[x] =  \l \int_0^1\int_0^1 d\s_1 d\s_2 (1-G(\s_1,\s_2))
\dot{x}^\mu(\s_1) \dot{x}^\mu(\s_2) \non 
\times\, \delta^{(d)}(x(\s_1)-x(\s_2))
W(C_{x(\s_1)x(\s_2)}) W(C_{x(\s_2)x(\s_1)}) 
\label{Jeps}
\eea
with the extra factor $(1-G(\s_1,\s_2))$ in the integrand which
recovers the principal-value integral over $\s_2$ as $\eps\ra0$.
For smooth contours the difference between~\rf{11} and \rf{Jeps} is
of order $\eps$ and vanishes as $\eps\ra0$.
Analogously, the regularized version of~\rf{Jeps}, which is given at $\eps=0$
by the r.h.s.\ of \eq{rLe}, reads
\bea
\lefteqn{J^{(G)}[x] =  \l \int_0^1\int_0^1 d\s_1 d\s_2 (1-G(\s_1,\s_2))
\dot{x}^\mu(\s_1) \dot{x}^\mu(\s_2)} 
\non &&\times \!
 \int_{r(0)=x(\s_1)}^{r(\Lambda^{-2})=x(\s_2)} {\cal D} r
\e^{-\fr 12 \int_0^{\Lambda^{-2}} d \tau \dot{r}^2(\tau)}
W(C_{x(\s_1)x(\s_2)}r_{x(\s_2)x(\s_1)}) 
W(C_{x(\s_2)x(\s_1)}r_{x(\s_1)x(\s_2)}) .~~~~~~~
\label{rJeps}
\eea

Thus, our strategy of the iterative solution of the loop equation is to 
represent it in the form
\be
W [x] = 1 - {1\over 2}\int_0^\infty dA \left\{\LA J^{(G)} [x+\sqrt{A}\xi ] 
\RA_{\xi}^{(G)}
- \LA J^{(G)} [\sqrt{A}\xi ] \RA_{\xi }^{(G)} \right\}
\label{4eps}                 
\ee
and to solve the last equation iteratively at finite $\eps$ after which to put
$\eps=0$.

\subsection{The order $\l$: gluon propagator}

Through zeroth
order in $\lambda$, $J^{(G)}[x]$ vanishes so that \eq{4eps} yields
\be
W_0[x] = 1.
\ee

This solution for $W_0[x]$ is to be substituted into \eq{Jeps} 
(or its regularized version~\rf{rJeps}) to determine $J^{(G)}[x]$
through order $\lambda $. We represent $J^{(G)}[x]$ to this order  in the form
\be
J_1^{(G)}[x] = \lambda \int_0^1\int_0^1 d\s_1 d\s_2 (1-G(\s_1,\s_2))
\dot{x}^\mu(\s_1) \dot{x}^\mu(\s_2) 
\int{d^{d}p_1\over (2\pi )^{d}} {d^{d}p_2\over (2\pi )^{d}}
\e ^{i\sum^{2}_{j=1} p_j x_j} D(p_1,p_2)
\label{16}
\ee
with
\be
D(p_1,p_2) = (2\pi )^{d} \delta ^{(d)}(p_1 + p_2)
\e^{-\frac{(p_1-p_2)^2}{8\L^2}}
\label{17}
\ee
for the regularized current~\rf{rJeps} by virtue of \eq{rpi0}.

Let us substitute the current~\rf{16} into \eq{4eps}.
After the shift $x(\s)\ra x(\s)+\sqrt{A} \xi(\s)$, as is prescribed by 
the r.h.s.\ of \eq{4eps}, there appears the terms of three types which
contain: 1) $\dot{x}^{\mu}(\s_1)\dot{x}^{\mu}(\s_2)$, 2)
$\dot{x}^{\mu }(\sigma _1) \dot{\xi}^{\mu }(\sigma_2)$  and
$\dot{\xi}^{\mu }(\sigma _1) \dot{x}^{\mu }(\sigma_2)$, 3)
$\dot{\xi}^{\mu }(\sigma _1) \dot{\xi}^{\mu }(\sigma_2)$. 
Each of them can be averaged over $\xi $ by
virtue of Eqs.~\rf{10}, \rf{15} and \rf{dot15}. 

For the first one, which contains
$\dot{x}^{\mu}(\s_1)\dot{x}^{\mu}(\s_2)$, one gets
\bea
W_1[x] = - {\lambda \over 2}\int^{\infty }_{0}dA 
\int_0^1\int_0^1 d\s_1 d\s_2 (1-G_{12})
\dot{x}^\mu(\s_1) \dot{x}^\mu(\s_2) \non \times
 \int {d^{d}p_1\over (2\pi )^{d}}{d^{d}p_2\over (2\pi)^{d}}
\e ^{i\sum^{2}_{j=1} p_j x_j - {A\over 2} \sum^{2}_{i,j=1}
 p_i G_{ij} p_j } D(p_1,p_2) \,.
\label{18}
\eea 
The factor $(1-G_{12})$ can be absorbed by introducing
\be
\alpha = \L^{-2}+ 2A(1-G_{12})
\ee 
which gives finally
\be
W_1[x] =  - {\lambda \over 4(2\pi)^{\frac d2}}
\int^{\infty }_{\L^{-2}}d\alpha 
\oint_C dx_1^\mu \oint_C dx_2^\mu \int {d^{d}p\over
(2\pi )^{d}}
\e ^{i p (x_1-x_2) - \frac \a2 p^2 } \,.
\label{orderlambda}
\ee 
This coincides with the perturbation-theory diagram 
of order $\lambda$ (depicted in Fig.~\ref{Fey}a) 
with the proper-time regularized gluon propagator.   
\message{Be patient -- some more figures}
\begin{figure}[tbp]
\unitlength=1.00mm
\begin{picture}(33.00,50.00)(5,85)
\thicklines
\bezier{80}(11.00,108.00)(11.00,119.00)(19.00,119.00)
\bezier{64}(19.00,119.00)(28.00,121.00)(32.00,126.00)
\bezier{108}(32.00,126.00)(39.00,133.00)(41.00,117.50)
\bezier{148}(21.00,100.00)(43.00,101.00)(41.00,117.50)
\bezier{80}(11.00,108.00)(11.00,100.00)(21.00,100.00)
\thinlines
\multiput(11.30,110.00)(4.00,0.00){7}{
  \bezier{28}(0.00,0.00)(1.00,1.00)(2.00,0.00)}
\multiput(13.30,110.00)(4.00,0.00){7}{
  \bezier{28}(0.00,0.00)(1.00,-1.00)(2.00,0.00)}
\bezier{28}(39.30,110.00)(39.90,110.50)(40.50,110.50)
\put(26.00,94.00){\makebox(0,0)[ct]{{\bf a)}}}
\end{picture}
\begin{picture}(33.00,50.00)(-19,85)
\thicklines
\bezier{80}(11.00,108.00)(11.00,119.00)(19.00,119.00)
\bezier{64}(19.00,119.00)(28.00,121.00)(32.00,126.00)
\bezier{108}(32.00,126.00)(39.00,133.00)(41.00,117.50)
\bezier{148}(21.00,100.00)(43.00,101.00)(41.00,117.50)
\bezier{80}(11.00,108.00)(11.00,100.00)(21.00,100.00)
\thinlines
\bezier{28}(12.00,114.50)(12.50,114.50)(13.00,115.00)
\multiput(13.00,115.00)(4.00,0.00){7}{
  \bezier{28}(0.00,0.00)(1.00,1.00)(2.00,0.00)}
\multiput(15.00,115.00)(4.00,0.00){7}{
  \bezier{28}(0.00,0.00)(1.00,-1.00)(2.00,0.00)}
\multiput(11.00,107.00)(4.00,0.00){7}{
  \bezier{28}(0.00,0.00)(1.00,-1.00)(2.00,0.00)}
\multiput(13.00,107.00)(4.00,0.00){7}{
  \bezier{28}(0.00,0.00)(1.00,1.00)(2.00,0.00)}
\put(26.00,94.00){\makebox(0,0)[ct]{{\bf b)}}}
\end{picture} 
\begin{picture}(33.00,50.00)(-42,85)
\thicklines
\bezier{80}(11.00,108.00)(11.00,119.00)(19.00,119.00)
\bezier{64}(19.00,119.00)(28.00,121.00)(32.00,126.00)
\bezier{108}(32.00,126.00)(39.00,133.00)(41.00,117.50)
\bezier{148}(21.00,100.00)(43.00,101.00)(41.00,117.50)
\bezier{80}(11.00,108.00)(11.00,100.00)(21.00,100.00)
\thinlines
\multiput(24.50,110.00)(4.00,0.00){4}{
  \bezier{28}(0.00,0.00)(1.00,1.00)(2.00,0.00)}
\multiput(26.50,110.00)(4.00,0.00){4}{
  \bezier{28}(0.00,0.00)(1.00,-1.00)(2.00,0.00)}
\multiput(24.50,110.00)(-2.00,-3.33){3}{
  \bezier{28}(0.00,0.00)(-1.333,-0.333)(-1.00,-1.665)}
\multiput(23.50,108.335)(-2.00,-3.33){3}{
  \bezier{28}(0.00,0.00)(0.333,-1.167)(-1.00,-1.665)}
\multiput(24.50,110.00)(-2.00,3.33){3}{
  \bezier{28}(0.00,0.00)(-1.333,0.333)(-1.00,1.665)}
\multiput(23.50,111.665)(-2.00,3.33){2}{
  \bezier{28}(0.00,0.00)(0.333,1.167)(-1.00,1.665)}
\bezier{28}(19.50,118.25)(19.60,118.80)(20.00,119.30)
\put(26.00,94.00){\makebox(0,0)[ct]{{\bf c)}}}
\end{picture} 
\caption[x]   {\hspace{0.2cm}\parbox[t]{13cm}
{\small
   Planar diagrams  for $W[x]$: a) of order $\lambda$ with gluon
   propagator, and of order $\lambda^2$ b) with two noninteracting 
   gluons and c) with the three-gluon vertex. The corresponding
   analytic expressions are given by Eqs.~\rf{orderlambda}, \rf{b)} and 
   \rf{cc)}, respectively.}}
\label{Fey}
\end{figure}
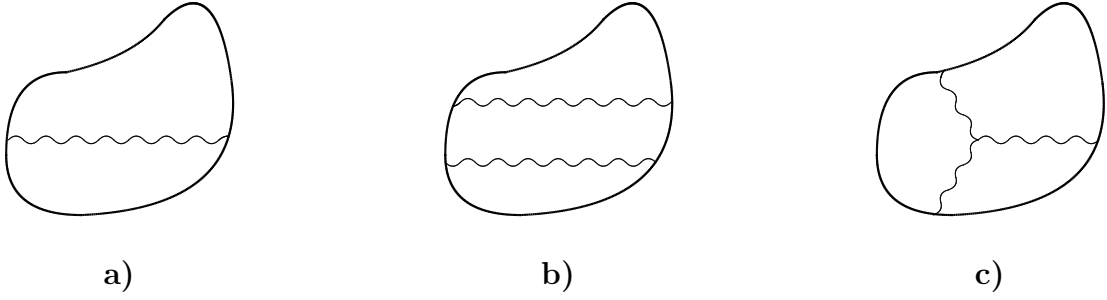

Let us show now that contributions of the remaining terms of the types 
2) and 3) vanish for smooth contours. 
The straightforward calculation of the terms of the type 2) 
at small $\eps$ yields
\be
W_1^{2)}[x] = \frac{\l}{12d(d-2)} \frac{\L^{d-2}}{(2\pi)^{\frac d2 - 2}}
\int_0^1 d \s \dot{x}^2(\s) \eps + {\cal O}(\eps^3)\,.
\label{2)}
\ee
Analogously, for the terms of the type 3) one gets 
\be
W_1^{3)}[x] = -\frac{\l}{12d(d-2)} \frac{\L^{d-2}}{(2\pi)^{\frac d2 - 2}}
\int_0^1 d \s \dot{x}^2(\s) \eps 
-\frac{\l\,\zeta(3)}{d} \frac{ \L^{d-2}}{(2\pi)^{\frac d2}}
\int_0^1 d \s \dot{x}^2(\s) \eps
+ {\cal O}(\eps^3)\,.
\label{3)}
\ee
Notice that the first term coincides with~\rf{2)} with the opposite 
sign. The coefficient in front of the second term is presented for $G$
given by \eq{Geps}. 

We see that \rf{2)} and \rf{3)}, which are not invariant under
a reparametrization, vanish 
for smooth contours when $|\dot{x}(\s)|\sim1$ provided  that $\eps $ tends to
zero at finite cutoff $\L$.  

It is worth mentioning that if one were use  alternatively
 the dimensional regularization  for which $d\neq 4$ while $\L=\infty$ in the
above formulas, the terms of the type 2) and 3) would vanish identically 
at any $\eps$.

\subsection{The order $\l^2$: three-gluon vertex}

Let us consider the next iteration of \eq{4eps}. The current
$J^{(G)}[x]$ through order $\lambda^2 $ reads
\be
J_2^{(G)}[x] = \lambda  
\int_0^1\int_0^1 d\s_1 d\s_2 (1-G_{12})
\dot{x}^\mu(\s_1) \dot{x}^\mu(\s_2) 
\delta ^{(d)}(x(\s_1)- x(\s_2)) \Big( W_1(C_{x_1 x_2}) + W_1(C_{x_2 x_1})\Big)  
\label{19}
\ee
or explicitly 
\bea
\lefteqn{J_2^{(G)}[x] = - \lambda^2  
\int_0^1 d\s_1\int_0^1 d\s_2 \int_0^1 d\s_3\int_0^1 d\s_4 (1-G_{12})
 (1-G_{34}) } \non&& \times\,
\dot{x}^\mu(\s_1) \dot{x}^\mu(\s_2)\dot{x}^\nu(\s_3) \dot{x}^\nu(\s_4)
\theta_c(1234) \int \prod _{i=1}^4
\left({d^d p_i\over (2\pi )^d}\right) \e^{i\sum^{4}_{j=1} p_j x_j}
D(p_1, \ldots, p_4) ~~~~~~~
\label{20} 
\eea 
where 
\be 
D(p_1,\ldots,p_4) = (2\pi )^{d} \delta 
^{(d)}(p_1 + p_2) (2\pi )^{d} \delta ^{(d)}(p_3 + p_4) \left\{{{1\over p_1^{2}} 
+ {1\over p_3^{2}}}\right\}\, .  
\label{21} 
\ee
For simplicity we restrict ourselves in this subsection by the 
dimensional regularization $d\neq4$.

To find $W_2[x]$, one substitutes $x(\s_i)\ra x(\s_i) + \sqrt{A} \xi (\s_i)$
into $J_2^{(G)}[x]$ after which a classification  
w.r.t.\ the number of $\dot{x}^\mu$ 
and $\dot{\xi}^\mu$ emerges. The simplest term is that with 4 $\dot x$'s
which is averaged over $\xi$ by using \eq{10}. 
The factors $(1-G_{12})$ and  $(1-G_{34})$ in \eq{20} are canceled by the 
contribution from the domains $|\s_1-\s_2|\la\eps$ and 
$|\s_3-\s_4|\la\eps$, respectively,
in full analogy with the above calculation of $W_1[x]$.
The result for  $\eps\ra0$ is
\bea
\lefteqn{W_2^{b)}[x] = \frac 12 \lambda ^2 
\int_0^1 d\s_1\int_0^1 d\s_2 \int_0^1 d\s_3\int_0^1 d\s_4 \,\theta_c(1234) 
\hspace{1.5cm} }  \non &&\times\,
\dot{x}^\mu(\s_1) \dot{x}^\mu(\s_2)\dot{x}^\nu(\s_3) \dot{x}^\nu(\s_4)
\int {d^d p_1\over (2\pi )^d} \int {d^d p_3\over (2\pi )^d}
\e^{ip_1(x_1-x_2)+ip_3(x_3-x_4)}  \frac{1}{p_1^2 p_3^2}~~~~
\label{b)}
\eea
which
coincides with the planar perturbation-theory diagram with 
two noninteracting gluons  depicted in Fig.~\ref{Fey}b.

It is a little bit more tedious 
to calculate the average of terms with 3 $\dot x$'s, 
1 $\dot \xi$ and with 2 $\dot x$'s, 2 $\dot \xi$'s which both contribute to the 
diagram with three-gluon vertex, depicted in Fig.~\ref{Fey}c.
There are four terms with 3 $\dot x$'s, 1 $\dot \xi$ which can be averaged 
over $\xi$ by virtue of \eq{15}. From the variety of terms with
 2 $\dot x$'s, 2 $\dot \xi$'s, which can be averaged over 
$\xi$ using \eq{dot15}, only two contribute to the diagram with the 
three-gluon vertex of  Fig.~\ref{Fey}c. They are associated with the
$\ddot G$ terms in the averages~\rf{dot15} 
of $\dot{\xi}(\s_1)\dot{\xi}(\s_4)$  and $\dot{\xi}(\s_2)\dot{\xi}(\s_3)$. 
Analogous terms in the averages of 
 $\dot{\xi}(\s_1)\dot{\xi}(\s_2)$  and $\dot{\xi}(\s_3)\dot{\xi}(\s_4)$
do not contain ordered contour integrals and vanish due to the general
arguments of Subsect.~\ref{ss4.4}. Combining the six terms, we get
\bea
W_2^{c)}[x] =  \lambda ^2 \int_0^\infty dA 
\int_0^1 d\s_1\int_0^1 d\s_2 \int_0^1 d\s_3\int_0^1 d\s_4 \,\theta_c(1234) 
(1-G_{12})(1-G_{34}) \hspace*{7mm}  \non \times
\int {d^d p_1\over (2\pi )^d} \int {d^d p_3\over (2\pi )^d}
\e^{ip_1(x_1-x_2)+ip_3(x_3-x_4)} 
\left( \frac{1}{p_1^2}+\frac{1}{p_3^2} \right)
\e^{-Ap_1^2-Ap_3^2+AG_{14}p_1p_3+AG_{23}p_1p_3}\,A \non \times
\Big\{\dot G_{41}ip^\mu_3 \dot{x}^\mu(\s_2)
\dot{x}^\nu(\s_3)\dot{x}^\nu(\s_1) +\dot G_{41}ip^\nu_1 \dot{x}^\nu(\s_3)
\dot{x}^\mu(\s_1)\dot{x}^\mu(\s_2) -
\dot G_{32}ip^\mu_3 \dot{x}^\mu(\s_1)
\dot{x}^\nu(\s_2)\dot{x}^\nu(\s_4) \hspace*{-.9cm}\non 
-\dot G_{32}ip^\nu_1 \dot{x}^\nu(\s_4)
\dot{x}^\mu(\s_1)\dot{x}^\mu(\s_2) 
-\ddot G_{14} \dot{x}^\mu(\s_2)\dot{x}^\mu(\s_3)
-\ddot G_{23} \dot{x}^\mu(\s_1)\dot{x}^\mu(\s_4)
\Big\}\,. \hspace{1.5cm}
\label{c)}
\eea

The three-gluon vertex appears as a result of doing the uncertainty
of the type $\eps\times \eps^{-1}$. It is evident for the first four terms
which involve $\dot G\sim 1/\eps$ while the essential
domain of integration is $|\s_1-\s_2|\la\eps$ or $|\s_3-\s_4|\la\eps$.
These four terms result, roughly speaking, in $2/3$ of the three-gluon 
vertex. The remaining $1/3$ results from the two
terms with $\ddot G$ in \eq{c)} which can be transformed, integrating 
 the ordered contour integral by parts, to a form with 3 $\dot x$'s. 
The resulting expression for $W^{c)}_2[x]$ as $\eps\ra0$ reads
\bea
\lefteqn{W_2^{c)}[x] =  \lambda ^2
\int_0^1 d\s_1\int_0^1 d\s_2 \int_0^1 d\s_3  \,\theta_c(123)
\dot{x}^\mu(\s_1)\dot{x}^\mu(\s_2) \dot{x}^\nu(\s_3) }\non &&
\times \int \prod _{i=1}^3
\left({d^d p_i\over (2\pi )^d}\right) \e^{i\sum^{3}_{j=1} p_j x_j}
(2\pi)^d \delta(p_1+p_2+p_3) i(p_1-p_2)_\nu \frac{1}{p_1^2 p_2^2 p_3^2}
\label{cc)}
\eea
with coincides with the diagram of Fig.~\ref{Fey}c.

The remaining terms with 2 $\dot x$'s, 2 $\dot \xi$'s as well as those
 with 1 $\dot x$, 3 $\dot \xi$'s and 
4 $\dot \xi$'s should reproduce the remaining diagrams of order $\l^2$
which have the form of the diagram of Fig.~\ref{Fey}a with the gluon and
ghost loop insertions into the propagator.

Generically, what is demonstrated in this section is how the solution of the functional
Laplace equation with an inhomogeneous term of Stokes' type brings back the functional 
of Stokes' type. The smearing procedure was crucial to get the three-gluon vertex by
doing the uncertainty $\eps\times \eps^{-1}$.


\newsection{Acknowledements}

I am grateful to Alexander Migdal for learning a lot from him during the great time of our
collaboration. 
I thank Tigran Shahbasyan for a useful correspondence about the emergence 
of the three-gluon vertex in early 1989. 
The two-day intersection with Marty Halpern at Berkeley in June 1988 which resulted
in Ref.~\cite{HM89} is unforgettable.
I thank Hikaru Kawai for useful discussions and his warm hospitality in Kyoto in 2005,
where Subsect.~\ref{s:regul} was added to the Notes.
This research was supported in part by the 
(no longer existing) International Science Foundation under grant MF-7000.




\begin{thebibliography}{33} 
\addtolength{\itemsep}{-6pt}
\small

\bibitem{MM79}
Yu.M. Makeenko and A.A. Migdal, 
{\it Exact equation for the loop average in multicolor QCD},
{Phys.~Lett.} {\bf 88B} (1979) 135.

\bibitem{MM81}
Yu.M. Makeenko and A.A. Migdal, 
{\it Quantum chromodynamics as dynamics of loops},
{Nucl.~Phys.} {\bf B188} (1981) 269.

\bibitem{Mig83}
A.A. Migdal, 
{\it Loop equations and $1/N$ expansion},
{Phys.~Rep.} {\bf 102} (1983) 199.

\bibitem{Lev51}
P. L\'{e}vy, {\it Probl\`{e}mes concrets d'analyse fonctionnelle}, Paris 1951.

\bibitem{Fel86}
M.N. Feller, {\it The L\' evy Laplacian,  166 Cambridge Tracts in Mathematics},
Cambridge Univ. Press (2005).

\bibitem{Mak02}
Y. Makeenko, {\it Methods of contemporary gauge theory},
Cambridge Univ. Press (2002), Chapter 12.

\bibitem{AK17}
    P.  Anderson and M. Kruczenski, {\it Loop equations and bootstrap methods in the lattice},
    Nucl.\ Phys.\ B {\bf 921} (2017) 702 [e-Print: 1612.08140 [hep-th]].

\bibitem{KZ23}
V. Kazakov and Z. Zheng, {\it Bootstrap for lattice Yang-Mills theory},
Phys.\ Rev. D {\bf 107} (2023)  L051501 [e-Print:  2203.11360 [hep-th]].

\bibitem{Li24}
Z. Li and S. Zhou, {\it Bootstrapping the Abelian lattice gauge theories,}
 JHEP 08 (2024) 154 [e-Print:  2404.17071 [hep-th]].

\bibitem{KZ24}
V. Kazakov and Z. Zheng,
{\it Bootstrap for finite N lattice Yang-Mills theory},
JHEP 03 (2025) 099 [e-Print: 2404.16925 [hep-th]].

\bibitem{Boo25}
Y. Guo, Z. Li, G. Yang, and G. Zhu,
{\it Bootstrapping SU(3) lattice Yang-Mills theory},
e-Print:  2502.14421 [hep-th].

\bibitem{Mig25}
A. Migdal, {\it Exact confining solution of the planar QCD loop equation via a matrix ensemble},
e-Print:  2507.05096 [hep-th].

\bibitem{turb}
A. Migdal,
{\it Statistical equilibrium of circulating fluids},
Phys.\ Rept.\ {\bf 1011} (2023) 1 [e-Print:  2209.12312 [physics.flu-dyn]].

\bibitem{Tav93}
J.N.~Tavares, {\it Chen integrals, generalized loops and loop calculus},
{preprint DF/IST 5.93} (May, 1993), hep-th/9305173.

\bibitem{Pol80}
A.M.~Polyakov, 
{\it Gauge fields as rings of glue},
{Nucl.~Phys.} {\bf B164} (1980) 172.

\bibitem{GN79}
J.L. Gervais and A. Neveu, 
{\it Local harmonicity of the Wilson loop integral in classical {Yang-Mills} theory},
{Nucl.~Phys.} {\bf B153} (1979) 445.

\bibitem{HM89}
M.B.~Halpern, Yu.M.~Makeenko, 
{\it Continuum regularized loop-space equation},
{Phys.\ Lett.} {\bf B218} (1989) 230.

\bibitem{PW81}
G. Parisi, Y.-S. Wu, 
{\it Perturbation theory without gauge fixing},
{Sci.~Sin.} {\bf 24} (1981) 483.

\bibitem{Zwa81}
D. Zwanziger,
{\it Covariant quantization of gauge fields without gribov ambiguity},
 {Nucl.~Phys.} {\bf B192} (1981) 259.

\bibitem{Mar81}
G. Marchesini, 
{\it A somment on the stochastic quantization: the loop equation of gauge theory 
as the equilibrium condition},
{Nucl.~Phys.} {\bf B191} (1981) 214.

\bibitem{BHST87}
Z.~Bern, M.B.~Halpern, L.~Sadun and C.~Taubes, 
{\it Continuum regularization of quantum field theory. 2. Gauge theory},
{Nucl.~Phys.} {\bf B284} (1987) 35.

\bibitem{Mak88}
Yu.M.~Makeenko, 
{\it Polygon discretization of the loop-space equation},
{Phys.~Lett.} {\bf B212} (1988) 221.

\bibitem{Mal98b}
J.~Maldacena,
{\it Wilson loops in large $N$ field theories},
Phys.\ Rev.\ Lett.\  {\bf 80} (1998) 4859
[{hep-th/9803002}].

\bibitem{RY98}
S.-J.~Rey and J.~Yee,
{\it Macroscopic strings as heavy quarks in large $N$ gauge theory and
anti-de Sitter supergravity,}
Eur.\ Phys.\ J.\ C {\bf 22} (2001) 379
[{hep-th/9803001}].

\bibitem{DGO99}
N.~Drukker, D.~J.~Gross and H.~Ooguri,
{\it Wilson loops and minimal surfaces},
Phys.\ Rev.\ D {\bf 60} (1999) 125006
[{hep-th/9904191}].

\bibitem{MOS06}
 Y. Makeenko, P. Olesen and 
G. W. Semenoff,
{\it Cusped SYM Wilson loop at two loops and beyond},
Nucl.\ Phys.\ {bf B748} (2006) 170 [hep-th/0602100].

\bibitem{FKKT97}
  M.~Fukuma, H.~Kawai, Y.~Kitazawa and A.~Tsuchiya,
  {\it String field theory from IIB matrix model,}
  Nucl.\ Phys.\ B {\bf 510} (1998) 158
  [{hep-th/9705128}].
  
\bibitem{IIB} 
H. Aoki, S. Iso, H. Kawai, Y. Kitazawa, and A. Tsuchiya,
{\it  IIB matrix model},
Prog.\ Theor.\ Phys.\ Suppl.\ {\bf 134} (1999) 47
  [hep-th/9908038].
  
\bibitem{Gat37}
R. G\^ ateaux, {\it Th\' eorie generale des fonctionnelles}, Gauthier-Villars, Paris (1937).


\bibitem{Mig86}
A.A. Migdal, 
{\it Momentum loop dynamics and random surfaces in QCD},
{Nucl.~Phys.} {\bf B265 [FS15]} (1986) 594.

\bibitem{BVM86}
M.A.~Bershadski, I.D.~Vaisburd and A.A.~Migdal,  
{\it Fourier functional transformations and loop equation},
{Sov.~J.~Nucl.~Phys.}  {\bf 43} (1986) 319.



\end{thebibliography}
\end{document}